%% file: DARS-Exp.tex
\documentclass[fleqn]{elsarticle}
\makeatletter
\def\ps@pprintTitle{%
	\let\@oddhead\@empty
	\let\@evenhead\@empty
	\def\@oddfoot{}%
	\let\@evenfoot\@oddfoot}
\makeatother
\usepackage{lineno}

\modulolinenumbers[5]

\usepackage{natbib}
\usepackage{amsmath,amsfonts,amssymb,epsfig,epstopdf,url,array}

\usepackage{amsthm}
\usepackage{color,soul}
\usepackage{algorithmic}
\usepackage{balance}
\usepackage[ruled]{algorithm2e}
\usepackage{stmaryrd}
\usepackage[table]{xcolor}
\usepackage{centernot}
\usepackage{tikz}
\usepackage{booktabs}
\usepackage{pdfpages}
\usepackage{graphicx}
\usepackage{subfigure}
\usepackage[font=small,skip=8pt]{caption}
\usepackage{tabularx}
\usepackage{comment}
\usepackage{multirow}
\usepackage{multicol}
\usepackage{bigstrut}
\usepackage{pdflscape}
\usepackage{rotating}
\usepackage{float}
\usepackage{lipsum}
\usepackage{enumitem}
\usetikzlibrary{plotmarks,shapes,arrows,chains,hobby,backgrounds,calc,trees}
\usepackage{makecell}
\usepackage{dblfloatfix}    
\usepackage{pdfcomment}
\usepackage{arydshln}
\usepackage{amssymb}
\usepackage{bm}
\usepackage{arydshln}
\usepackage{pgfplots}
\usepackage{pgfplotstable}
\pgfplotsset{compat=newest}
\usepackage{lipsum}                     
\usepackage{xargs}                      
\usepackage[colorinlistoftodos,prependcaption,textsize=tiny]{todonotes}
\newcommandx{\unsure}[2][1=]{\todo[linecolor=red,backgroundcolor=red!25,bordercolor=red,#1]{#2}}
\newcommandx{\change}[2][1=]{\todo[linecolor=blue,backgroundcolor=blue!25,bordercolor=blue,#1]{#2}}
\newcommandx{\info}[2][1=]{\todo[linecolor=OliveGreen,backgroundcolor=OliveGreen!25,bordercolor=OliveGreen,#1]{#2}}
\newcommandx{\improvement}[2][1=]{\todo[linecolor=orange,backgroundcolor=orange!25,bordercolor=orange,#1]{#2}}
\newcommandx{\hiddencomment}[2][1=]{\todo[disable,#1]{#2}}

\definecolor{lightgray}{gray}{0.9}
\theoremstyle{definition}
\newtheorem{exmp}{Example}
\theoremstyle{definition}
\newtheorem{mydef}{Proposition}
\theoremstyle{definition}

\newcommand*{\Perm}[2]{{}^{#1}\!P_{#2}}%

\biboptions{authoryear}
\biboptions{authoryear}

\definecolor{header}{rgb}{0.0,0.0,0.0}
\definecolor{myblue}{rgb}{0.5,0.5,0.5}

\begin{document}

\begin{frontmatter}



\title{Dependency-Aware Software Requirements Selection using Fuzzy Graphs and Integer Programming}




\author{Davoud Mougouei *}
\address{School of Computing and IT, University of Wollongong, NSW, Australia \\dmougouei@gmail.com}

\author{David M. W. Powers}
\address{College of Science and Engineering, Flinders University, SA, Australia \\David.Powers@flinders.edu.au}


\begin{abstract}
\input{abstract}
\end{abstract}

\begin{keyword}
Fuzzy; Integer Programming; Value; Dependencies; Software
\end{keyword}

\end{frontmatter}

\input{introduction}
\input{related}
\input{identification}
\input{modeling}

\input{selection}
\input{selection_ov}

\input{selection_ilp}
\input{case}
\input{case_description}
\input{case_identification}
\input{case_selection}
\input{scalability}
\smaller
\bibliographystyle{apalike}
\bibliography{ref}

\end{document}

%% file: abstract.tex
Software requirements selection aims to find an optimal subset of the requirements with the highest \textit{value} while respecting the project constraints. But the value of a requirement may depend on the presence or absence of other requirements in the optimal subset. Such \textit{Value Dependencies}, however, are imprecise and hard to capture. In this paper, we propose a method based on integer programming and fuzzy graphs to account for value dependencies and their imprecision in software requirements selection. The proposed method, referred to as \textit{Dependency-Aware Software Requirements Selection} (DARS), is comprised of three components: (i) an automated technique for the identification of value dependencies from user preferences, (ii) a modeling technique based on fuzzy graphs that allows for capturing the imprecision of value dependencies, and (iii) an \textit{Integer Linear Programming} (ILP) model that takes into account user preferences and value dependencies identified from those preferences to reduce the risk of value loss in software projects. Our work is verified by studying a real-world software project. The results show that our proposed method reduces the value loss in software projects and is scalable to large requirement sets. 

%% file: introduction.tex
\section{Introduction}
\label{sec_introduction}

Software requirement selection, also known as \textit{Release Planning}, is to find a subset of requirements that delivers the highest value for a release of software while respecting the project constraints, e.g. budget~\citep{Zhang:2018:ESM:3208361.3196831,zhang2018empirical,deMeloFranca_2018,bagnall_next_2001,franch2016software}. Selecting (ignoring) a requirement, however, may influence the values of other requirements~\citep{aydemir2018next,mougouei2016factoring,mougouei2017dasrp,mougouei2020dependency,Zhang_RIM_2013,Robinson_RIM_2003}; it is important to consider value dependencies in requirement selection~\citep{carlshamre_industrial_2001,li_integrated_2010,zhang_investigating_2014,karlsson_improved_1997,mougouei2019dependency}. This is further emphasized by the fact that value dependencies widely exist in software projects~\cite{carlshamre_industrial_2001,carlshamre_release_2002,pitangueira2015software}. Moreover, as observed by Carlshamre et al.~\citep{carlshamre_industrial_2001}, requirement dependencies and value dependencies in particular are \emph{fuzzy}~\citep{carlshamre_industrial_2001} as the strengths of those dependencies are imprecise and vary ~\citep{dahlstedt2005requirements,ngo_wicked_2008,ngo2005measuring,carlshamre_industrial_2001} from large to insignificant~\citep{wang_simulation_2012}. 

Although the need for considering value dependencies and their strengths was observed as early as in 2001~\citep{carlshamre_industrial_2001}, existing requirements selection methods have mainly ignored value dependencies by simply optimizing either the \textit{Accumulated Value} (AV)~\citep{baker_search_2006,li_integrated_2010,boschetti_lagrangian_2014,araujo2016architecture,Greer_evolutionary_2004} or the \textit{Expected Value} (EV) of the selected requirements~\citep{pitangueira2017minimizing,li2016value,li2014robust}. Some methods have attempted considering value dependencies by manually estimating the values of requirement subsets, that may require up to $2^n$ comparisons for $n$ requirements~\citep{van_den_akker_flexible_2005}. When estimations are limited to pairs, still $O(n^2)$ estimations are needed ~\citep{li_integrated_2010,sagrado_multi_objective_2013,Zhang_RIM_2013}. Such complexity severely impacts the practicality of these methods, not to mention the issues around the accuracy of manual estimations. 

Finally, requirements selection methods based on manual estimations of values of requirement subsets do not capture the direction of an influence. In other words, these methods do not distinguish among three scenarios: (i) requirement $r_i$ influences the value of requirement $r_j$ and not the other way, (ii) $r_j$ influences the value of $r_i$ and not the other way, and (iii) both $r_i$ and $r_j$ influence the values of each other but to different extents. To effectively consider value dependencies in software requirements selection, we have proposed a method based on fuzzy graphs and integer programming~\citep{gkioulekas2019piecewise,tavana2015fuzzy} with three major components:

(i) \textit{Identification of value dependencies}. We have contributed an automated technique that uses \text{Eells} measure of casual strength~\citep{eells1991probabilistic} to extract value dependencies from significant causal relations among user preferences. \textit{Odds Ratio}~\citep{li2016observational} is used to specify the significance of such relations. We have further, demonstrated the use of a Latent Multivariate Gaussian model~\citep{macke2009generating} to generate samples of user preferences when needed~\citep{macke2009generating}. 

(ii) \textit{Modeling value dependencies}. We have demonstrated the use of fuzzy graphs~\citep{zahedi2016swarm,rosenfeld_fuzzygraph_1975} and their algebraic structure~\citep{kalampakas_fuzzy_2013} for modeling strengths and qualities of value dependencies and capturing the imprecision of those dependencies. On this basis, value dependencies and their strengths are modeled by fuzzy relations~\citep{carlshamre_industrial_2001,ngo_fuzzy_2005_structural,ngo2005measuring,liu_imprecise_1996} and their fuzzy membership functions respectively. 

(iii) \textit{Considering value dependencies in requirements selection}. At the heart of DARS is an integer linear programming (ILP) model, which maximizes the \textit{Overall Value} (OV) of a selected subset of the requirements, where user preferences and the value dependencies identified from those preferences are considered. The model further, considers structural and semantic dependencies (e.g. Requires and Conflicts-With) among software requirements by formulating them as the precedence constraints of the optimization model. 

We have demonstrated practicality and scalability of DARS by studying a real-world software. Our results show that (a) compared to existing requirements selection methods, that ignore value dependencies, DARS provides higher overall value by mitigating the risk of value loss caused by ignoring (selecting) requirements with positive (negative) influences on the values of the selected requirements, (b) maximizing the accumulated value or the estimated value of a software conflicts with maximizing its overall value, and (c) DARS is scalable to datasets with large number of requirements for different levels of value dependencies and precedence dependencies among requirements. This is demonstrated by simulating different scenarios for datasets of up to $3000$ requirements.  

%% file: related.tex
\section{Background and Related Work}
\label{sec_related}

It is widely recognized that software requirements influence the values of each other~\citep{Zhang_RIM_2013,brasil_multiobjective_2012,Robinson_RIM_2003,dahlstedt2005requirements}. Such influences are described in the literature as value dependencies~\citep{carlshamre_industrial_2001,li_integrated_2010,zhang_investigating_2014,karlsson_improved_1997}. Value dependencies are fuzzy relations~\citep{carlshamre_industrial_2001} with varying strengths (e.g. weak, moderate, strong) and qualities (positive or negative) which are imprecise, and hard to specify~\citep{carlshamre_industrial_2001,mougouei2016factoring}. Requirement selection methods hence, should consider qualities and strengths of explicit and implicit value dependencies among software requirements.

Moreover, \textit{Precedence Dependencies} such as \textit{Requires}~\citep{dahlstedt2005requirements}, \textit{Conflicts-With}~\citep{k_process_centered_1996}, \textit{AND}, and \textit{OR} also have value implications. For instance, a requirement $r_i$ requires (conflicts-with) $r_j$ means that $r_i$ cannot give any value if $r_j$ is ignored (selected). Hence, it is also important to consider value implications of precedence dependencies in software requirements selection. On this basis, we have characterized properties (P1)-(P7) for requirements selection methods in relation to how they treat value dependencies. 

Table~\ref{table_lr} categorizes existing requirements selection methods into four groups based on (P1)-(P7).   

\begin{itemize}
	\itemsep0em 
	\item[(P1)] Considering explicit value dependencies.
	\item[(P2)] Considering implicit value dependencies.
	\item[(P3)] Considering qualities (positive or negative) of value dependencies.
	\item[(P4)] Considering strengths of value dependencies.
	\item[(P5)] Considering directions of value dependencies.
	\item[(P6)] Considering precedence dependencies and their value implications.
	\item[(P7)] Relying on manual estimations of values of requirement subsets.      
\end{itemize}

\makeatletter
\def\Cline#1#2{\@Cline#1#2\@nil}
\def\@Cline#1-#2#3\@nil{%
	\omit
	\@multicnt#1%
	\advance\@multispan\m@ne
	\ifnum\@multicnt=\@ne\@firstofone{&\omit}\fi
	\@multicnt#2%
	\advance\@multicnt-#1%
	\advance\@multispan\@ne
	\leaders\hrule\@height#3\hfill
	\cr}
\makeatother


\begin{table}[!htb]
	\setlength\arrayrulewidth{1.5pt}
	\caption{Considering aspects of value dependencies in existing works.}
	\label{table_lr}
	\centering
	\input{table_lr4}
\end{table}
\vspace{0.2cm}

\subsection{Binary Knapsack Methods}
\label{sec_related_bk}

The first group of selection methods (Table~\ref{table_lr}), i.e. \textit{Binary Knapsack} (BK) methods, are solely based on the classical formulation of binary knapsack problem~\citep{harman_exact_2014,carlshamre_industrial_2001} as given by (\ref{eq_pcbk})-(\ref{eq_pcbk_c2}). Let~$R=\{r_1,...,r_n\}$ be a set of identified requirements, where $\forall r_i\in R$ ($1 \leq i \leq n$), $v_i$ and $c_i$ in (\ref{eq_pcbk})-(\ref{eq_pcbk_c2}) denote the value and the cost of $r_i$ respectively. Also, $b$ in (\ref{eq_pcbk_c1}) denotes the available budget. A decision variable $x_i$ specifies whether requirement $r_i$ is selected ($x_i=1$) or not ($x_i=0$). The objective of \textit{BK} methods as given by (\ref{eq_pcbk}) is to find a subset of $R$ that maximizes the accumulated value of the selected requirements $(\sum_{i=1}^{n} v_i  x_i)$ while entirely ignoring value dependencies as well as precedence dependencies among the requirements~\citep{karlsson_optimizing_1997,jung_optimizing_1998,zhang_multi_objective_2007}. 

\begin{align}
\label{eq_pcbk}
& \text{Maximize} \sum_{i=1}^{n} v_i x_i \\
\label{eq_pcbk_c1}
& \text{Subject to} \sum_{i=1}^{n} c_i  x_i \leq b \\
\label{eq_pcbk_c2}
& x_i \in \{0,1\},\quad i = 1,...,n
\end{align}

\subsection{Precedence-Constrained Binary Knapsack Methods}
\label{sec_related_pcbk}

The \textit{Precedence-Constrained Binary Knapsack} (PCBK) methods, enhance the BK methods by adding (\ref{eq_pcbk_c3}) to the optimization model of BK methods to consider precedence dependencies (\textit{requires}~\citep{dahlstedt2005requirements} and \textit{conflicts-with}~\citep{k_process_centered_1996}) and their value implications. A positive (negative) dependency from a requirement $r_j$ to $r_k$ is denoted by $x_j\le x_k$ ($x_j\le 1-x_k$) in (\ref{eq_pcbk_c3}). Also, decision variable $x_i$ denotes whether a requirement $r_i$ is selected ($x_i=1$) or not. 

\begin{align}
\label{eq_pcbk_c3}
& \begin{cases}
x_j \le x_k  & \text{if $r_j$ positively depends on $r_k$} \\
x_j \le 1-x_k& \text{if $r_j$ negatively depends on $r_k$},\quad j\neq k= 1,...,n \\
\end{cases}
\end{align}

\vspace{0.1cm}
\subsection{Increase-Decrease Methods}
\label{sec_related_pcbk_CS}

The third group of requirement selection methods i.e. \textit{Increase-Decrease} methods consider value dependencies among requirements through estimating the amount of the increased (decreased) values resulted by selecting different subsets of requirements. The optimization model for an increases-Decreases technique proposed by~\citep{van_den_akker_flexible_2005} is given in (\ref{eq_pcbk-CS})-(\ref{eq_pcbk-CS_c3}). For a subset $ s_j \in S:\{s_1,...,s_m\}, m\leq2^n$, with $n_j$ requirements, the difference between the estimated value of $s_j$ ($w_j$) and the accumulated value of the requirements in $s_j$ ($\sum_{r_k\in s_j}v_k$) is considered when computing the value of the selected requirements. $y_j$ in (\ref{eq_pcbk-CS}) specifies whether a subset $s_j$ is realized ($y_j=1$) or not ($y_j=0$). Also, constraint~(\ref{eq_pcbk-CS_c2}) ensures that $y_j=1$ only if $\forall r_k \in s_j, x_k=1$.

\begin{align}
\label{eq_pcbk-CS}
&\text{Maximize} \sum_{i=1}^{n} v_i x_i + \sum_{j=1}^{m} (w_j-\sum_{r_k\in s_j}v_k)\text{ } y_j\\
\label{eq_pcbk-CS_c1}
&\text{Subject to }  n_j y_j \leq \sum_{r_k \in s_j} x_k \\
\label{eq_pcbk-CS_c2}
&\sum_{i=1}^{n} c_i  x_i \leq b \\
\label{eq_pcbk-CS_c3}
& x_i,y_j \in \{0,1\},\quad i = 1,...,n,\quad j = 1,...,m
\end{align}

Increase-Decrease methods are complex and prone to human error as they rely on manual estimations for requirement subsets~\citep{mougouei2016factoring}. For $n$ requirements, these estimations are at least as complex as $O(n^2)$, when only values of pairs are estimated~\citep{li_integrated_2010,sagrado_multi_objective_2013,Zhang_RIM_2013}, and can get as complex as $O(2^n)$ in worst case.  


\begin{align}
\label{eq_pcbk-Others}
&\text{Maximize } \sum_{i=1}^{n} v_i x_i + \sum_{i=1}^{n}\sum_{j=1}^{n} x_i x_j w_{i,j}\\
\label{eq_pcbk-Others_c1}
&\text{Subject to} \sum_{i=1}^{n} c_i  x_i \leq b \\
\label{eq_pcbk-Others_c2}
& y_{ij} \leq x_{i}\\
\label{eq_pcbk-Others_c3}
& y_{ij} \leq x_{j}\\
\label{eq_pcbk-Others_c4}
& y_{i,j} \geq x_{i} + x_{j} - 1\\
\label{eq_pcbk-Others_c5}
& x_i,y_{i,j} \in \{0,1\} \quad i,j = 1,...,n
\end{align}

Moreover, relying on pairwise estimations results in ignoring implicit value dependencies as the direction of dependencies are not specified. For instance, consider requirements $R:\{r_1,r_2,r_3\}$ with positive value dependencies from $r_1$ to $r_2$ and from $r_2$ to $r_3$. An implicit positive value dependency from $r_1$ to $r_3$ can be inferred. An increase-Decrease model, however, fails to capture this even if pairwise estimations identify the value of $r_1$ and $r_2$ ($r_2$ and $r_3$) as a pair is higher than the accumulated value of $r_1$ and $r_2$ ($r_2$ and $r_3$). If no explicit value dependency is found between $r_1$ and $r_3$ hence the influence of $r_3$ on the value of $r_1$ will be ignored.

\subsection{Stochastic Binary Knapsack Methods}
\label{sec_related_sbk}

\textit{Stochastic Binary Knapsack} (SBK) requirements selection methods maximize the expected value of a requirement subset based on the formulation of stochastic knapsack problem~\citep{henig1990risk} as given by~(\ref{eq_sbk}). In this equation, $E(v_i)$ denotes the expected value of a requirement $r_i$. The work~\citep{pitangueira2017minimizing} for instance, optimizes the expected value of a software at different risk levels, where risk is formulated in terms of summation of covariances of values of requirements as given by~(\ref{eq_sbk_c1}), where $\sigma_{i,j}$ specifies the covariance of $v_i$ and $v_j$ and $l$ specifies the risk level.

\begin{align}
\label{eq_sbk}
\text{Maximize } & \sum_{i=1}^{n} x_i E(v_i)\\
\label{eq_sbk_c1}
\text{Subject to} & \sum_{i=1}^{n} \sum_{j=1}^{n} x_ix_j \sigma_{i,j} \leq l\\ 
\label{eq_sbk_c2}
& \sum_{i=1}^{n} x_i c_i \leq b\\
\label{eq_sbk_c3}
&x_i \in \{0,1\}\quad \quad \quad i = 1,...,n
\end{align}


One may suggest that the covariance of values of requirements (risk) in~(\ref{eq_sbk_c1}) may somehow capture value dependencies. But, there are four major problems with this. First, covariance is a measure of correlation and does not capture causality. In other words, by using covariance one is assuming that all value dependencies are bidirectional and the strengths of dependencies in either direction are equal. Such an assumption, however, may not be realistic. Second, optimization models based on covariance can only capture linear relations ignoring non-linear relations even if they are significant. 

Also, limiting or minimizing the risk may contradict with choosing requirements that positively influence each other as SBK methods avoid choosing positively correlated (with respect to value) requirements. On the contrary, SBK methods tend to choose negatively correlated requirements or independent ones as such combinations satisfy (\ref{eq_sbk_c1}). In other words, risk, which is defined based on covariances, and value dependencies, which are, by nature, causal relations, are different and cannot be used interchangeably. 

It is also worth mentioning that Value (of the requirement/software) is a broad concept and has several manifestations as discussed in~\cite{Mougouei:2018:OHV:3236024.3264843,perera2019towards,hussain2018integrating,perera2019study}. In this paper, however, we only focus on the monetary interpretation of Value. 

%% file: table_lr4.tex
\Huge
\resizebox {1\textwidth }{!}{
	\begin{tabular}{|l|l|l|l|l|l|l|l|l|l|}
		\toprule[1.5 pt]
		\textbf{\cellcolor{header}\textcolor{white}{Technique}} &
		\textbf{\cellcolor{header}\textcolor{white}{Employed by}} &
		\textbf{\cellcolor{header}\textcolor{white}{P1}} &
		\textbf{\cellcolor{header}\textcolor{white}{P2}} &
		\textbf{\cellcolor{header}\textcolor{white}{P3}} &
		\textbf{\cellcolor{header}\textcolor{white}{P4}} &
		\textbf{\cellcolor{header}\textcolor{white}{P5}} &
		\textbf{\cellcolor{header}\textcolor{white}{P6}} &
		\textbf{\cellcolor{header}\textcolor{white}{P7}} 

		\bigstrut\\
		\hline
		BK &
		\begin{tabular}[t]{@{}c@{}} \cite{karlsson_optimizing_1997,jung_optimizing_1998,zhang_multi_objective_2007,baker_search_2006,finkelstein2009search,zhang2011comparing,del2010ant,kumari2012software} \phantom{sssssssssssss} \end{tabular} &
		NO &
		NO &
		NO &
		NO &
		NO &
		NO &
		NO
		\bigstrut\\
		\hline
		\multirow{4}[9]{*}{PCBK} &
		\begin{tabular}[t]{@{}l@{}} \cite{veerapen2015integer,brasil_multiobjective_2012,sagrado_multi_objective_2013,bagnall_next_2001,greer_software_2004,boschetti_lagrangian_2014,ruhe_quantitative_2003,van2011quantitative,zhang2010search,tonella2010using,freitas2011software}\\ \cite{colares_new_2009,saliu2007bi,saliu2005supporting,jiang2010hybrid,van2005determination,ngo2009optimized,chen2013ant,del2011requirements,araujo2016architecture,colares2009new,pitangueira2016risk}\\\cite{van2008software,ngo_wicked_2008,tonella2013interactive,xuan2012solving,saliu2005software}\end{tabular} &
		NO  &
		NO  &
		NO  &
		NO  &
		NO  &
		YES  &
		NO  
		\bigstrut\\
		\hline
		\multicolumn{1}{|l|}{\multirow{2}[4]{*}{Increase-Decrease}} &
		\cite{li_integrated_2010,sagrado_multi_objective_2013,Zhang_RIM_2013} (subsets of size 2) &
		YES &
		NO  &
		YES &
		NO  &
		NO  &
		YES &
		YES 
		\bigstrut\\
		\cline{2-10}\multicolumn{1}{|l|}{} &
		\cite{van_den_akker_flexible_2005} (subsets of any size) &
		YES &
		YES &
		NO  &
		NO  &
		NO  &
		YES &
		YES 
		\bigstrut\\
		\hline
		SBK &
		\begin{tabular}[t]{@{}c@{}} \cite{pitangueira2017minimizing,li2016value,li2014robust} \phantom{sssssssssssss} \end{tabular} &
		NO &
		NO &
		NO &
		NO &
		NO &
		NO &
		YES
		\bigstrut\\
		\hline
	\end{tabular}%
}

%% file: identification.tex
\section{Identification of Value Dependencies}
\label{ch_dars_identification}
\hypertarget{ch_dars_identification}{ }    
This section presents an automated technique for identification of value dependencies based on causal relations among user preferences for requirements. We use the widely adopted Eells measure~\citep{eells1991probabilistic} of causal strength and the \textit{Odds Ratio}~\citep{li2016observational} to identify the qualities and strengths of significant causal relations among requirements. A fuzzy membership function will then be used to estimate the strengths and qualities of value dependencies based on identified causal relations. Identified value dependencies will be used to identify implicit dependencies among requirements using the algebraic structure of fuzzy graphs.  

\subsection{Gathering User Preferences}
\label{ch_dars_identification_gathering}

User preferences can be gathered in different ways~\citep{leung2011probabilistic,holland2003preference,sayyad2013value} depending on the nature of the release. For a new software product, preferences may be gathered by conventional market research techniques such as conducting surveys and mining user reviews in social media and online stores~\citep{villarroel2016release}. User preferences may also be gathered by studying similar software and sales records. 

When sales/usage records for the requirements of a software product are available, say from earlier versions, such information can be combined with market research results to estimate user preferences for a newer version of the software. This is particularly suitable for reengineering software or releasing different configurations of a software product line. We capture user preferences by a \textit{Preference Matrix} as given by Definition~\ref{def_pm}.

\begin{figure*}
    \begin{center}
        \includegraphics[scale=1]{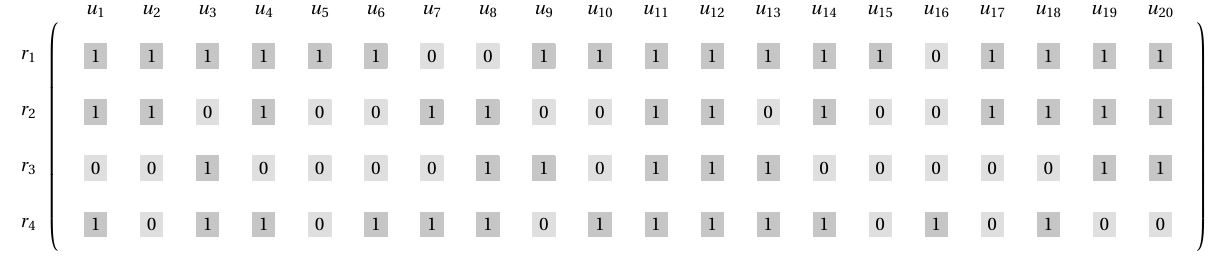}
    \end{center}
    \vspace{-0.25cm}
    \caption{A sample preference matrix $M_{4\times20}$.}
    \label{fig_ch_dars_pm}
\end{figure*}

\begin{mydef}
    \label{def_pm}
    \textit{Preference Matrix}. Let $R=\{r_1,...,r_n\}$ be a requirement set and $U=\{u_1,...,u_k\}$ be the list of users whose preference are gathered. A preference matrix $M$ is a binary ($ 0/1 $) matrix of size $n \times k$ where $n$ and $k$ denote the number of requirements and the number of users respectively. Each element $m_{i,j}$ specifies whether a user $u_i$ has preferred a requirement $r_j$ ($m_{i,j}=1$) or not ($m_{i,j}=0$). A sample preference matrix $M_{4\times 20}$ is shown in Figure~\ref{fig_ch_dars_pm}.
\end{mydef}

\vspace{0.5cm}
\subsection{Resampling}
\label{ch_dars_identification_resampling}

Resampling may be required to generate samples of user preferences based on the estimated distribution of the original data (collected user preferences) to enhance the accuracy of Eells measure. This is particularly useful when conducting comprehensive market research is not practical. 

\begin{figure}[!htb]
    \centering
    \centerline{\includegraphics[scale=0.8]{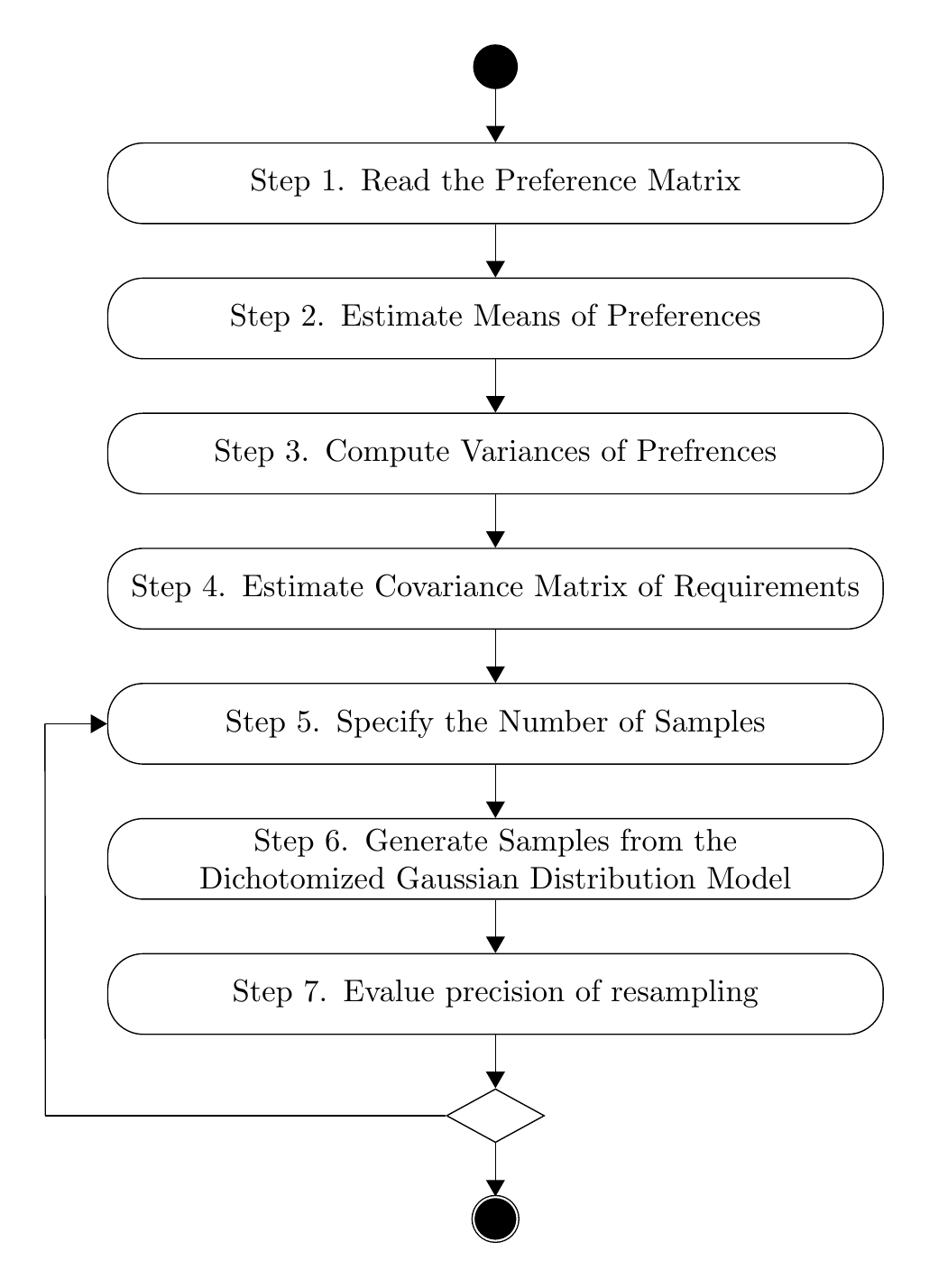}}
    \vspace{-0.0cm}
    \caption{Steps for generating samples from user preferences.}
    \label{fig_ch_dars_resampling}
\end{figure}

We use a resampling technique introduced by~\citep{kroese2014statistical} to generate larger samples of collected user preferences using a Latent Multivariate Gaussian model. The process as given in Figure~\ref{fig_ch_dars_resampling} starts with reading the preference matrix of users (Step 1) and continues with estimating the means (Step 2) and variances of user preferences (Step 3) for each requirement. Then the covariance matrix of the requirements will be computed (Step 4) to be used for generating new samples. Thereafter, the number of samples will be specified (Step 5) and samples will be generated based on the Dichotomized Gaussian Distribution model discussed in~\citep{kroese2014statistical}, Step 6. 

The precision of the employed resampling technique (Figure~\ref{fig_ch_dars_resampling}) can be evaluated (Step 7) by comparing the means and covariance matrix of the generated samples against the covariance matrix of the initial samples gathered from users. Steps 1 to 7 may be repeated for larger numbers of samples until the means and covariance matrix of the resampled data and those of the initial sample converge.    

Macke's technique has proved to be computationally efficient and feasible for a large number of variables (software requirements). The entropy of the Latent Multivariate Gaussian model is near theoretical maximum for a wide range of parameters~\citep{macke2009generating}. 

\subsection{Extracting Causal Relations among User Preferences}
\label{ch_dars_identification_relations}

User preferences for a requirement may positively or negatively influence the preferences of the users for other requirements. Such causal relations can be identified using measures of causal strength~\citep{Halpern01062015,pearl2009causality,janzing2013quantifying}. Causal relations among user preferences can then be used to specify the strengths and qualities of value dependencies among requirements as values of software requirements are determined by user preferences for those requirements.

As such, we have adopted one of the most widely used measures of causal strength, referred to as Eells measure~\citep{eells1991probabilistic}, to estimate the strengths and qualities of explicit value dependencies among software requirements as given by (\ref{Eq_ch_dars_Eells}). The sign (magnitude) of $\eta_{i,j}$ specifies the quality (strength) of a value dependency from a requirement $r_i$ to $r_j$, where selecting (ignoring) $r_j$ may influence, either positively or negatively, the value of $r_i$. 

\begin{align}
\label{Eq_ch_dars_Eells}
& \eta_{i,j}= p(r_i|r_j) - p(r_i|\bar{r_j}) ,\phantom{s}\eta_{i,j} \in [-1,1]
\end{align}

For a pair of requirements ($r_i,r_j$), Eells measure captures both positive and negative value dependencies from $r_i$ to $r_j$ by subtracting the conditional probability $p(r_i|\bar{r_j})$ from $p(r_i|r_j)$, where conditional probabilities $p(r_i|\bar{r_j})$ and $p(r_i|r_j)$ denote strengths of positive and negative causal relations from $r_i$ to $r_j$ respectively, that is selecting $r_i$ may increase or decrease the value of $r_j$.

\begin{figure}[!htb]
    \begin{center}
        \subfigure[$P_{4 \times 4}$]{%
            \label{fig_ch_dars_eta_p}
            \includegraphics[scale=0.3]{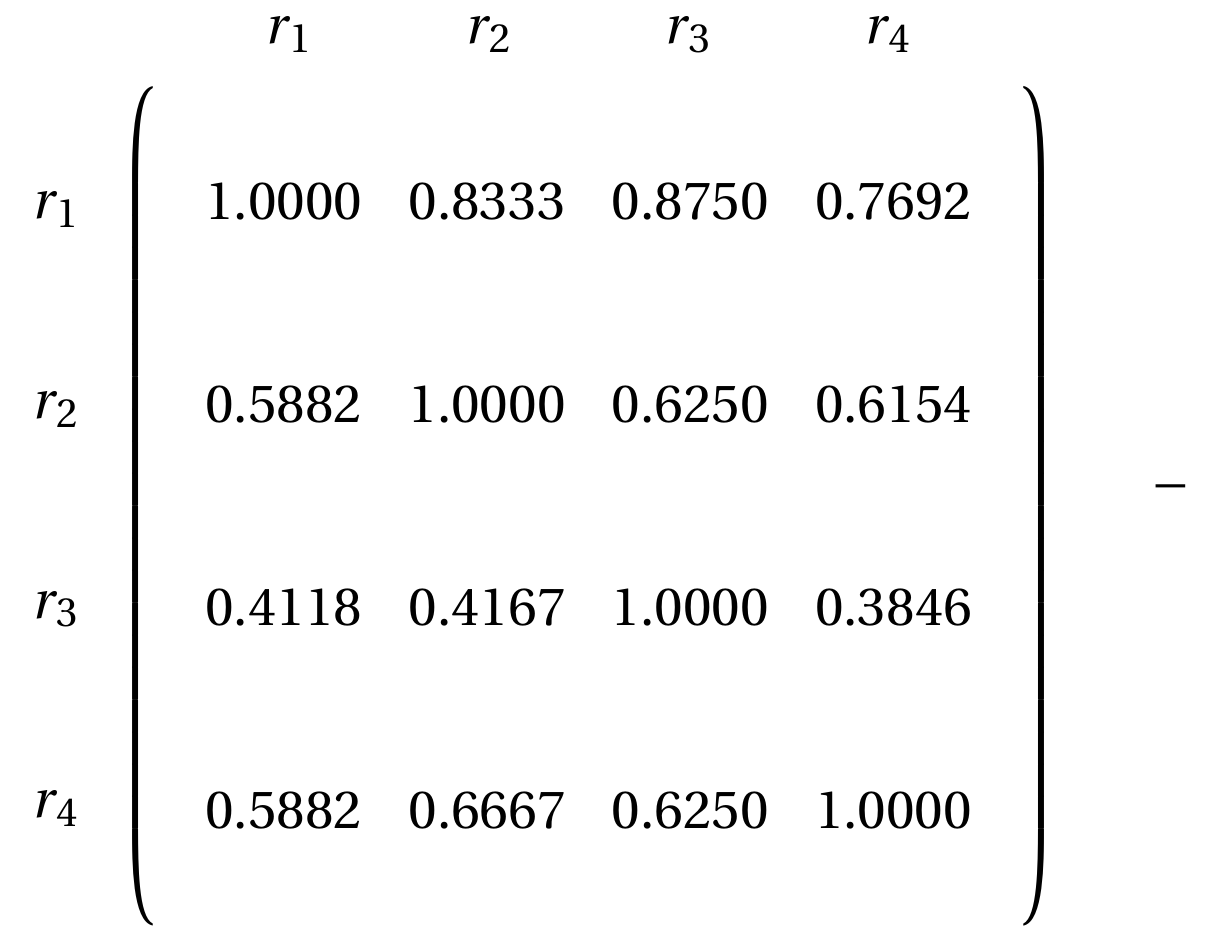}
        }
        \subfigure[$\bar{P}_{4 \times 4}$]{%
            \label{fig_ch_dars_eta_p_bar}
            \includegraphics[scale=0.3]{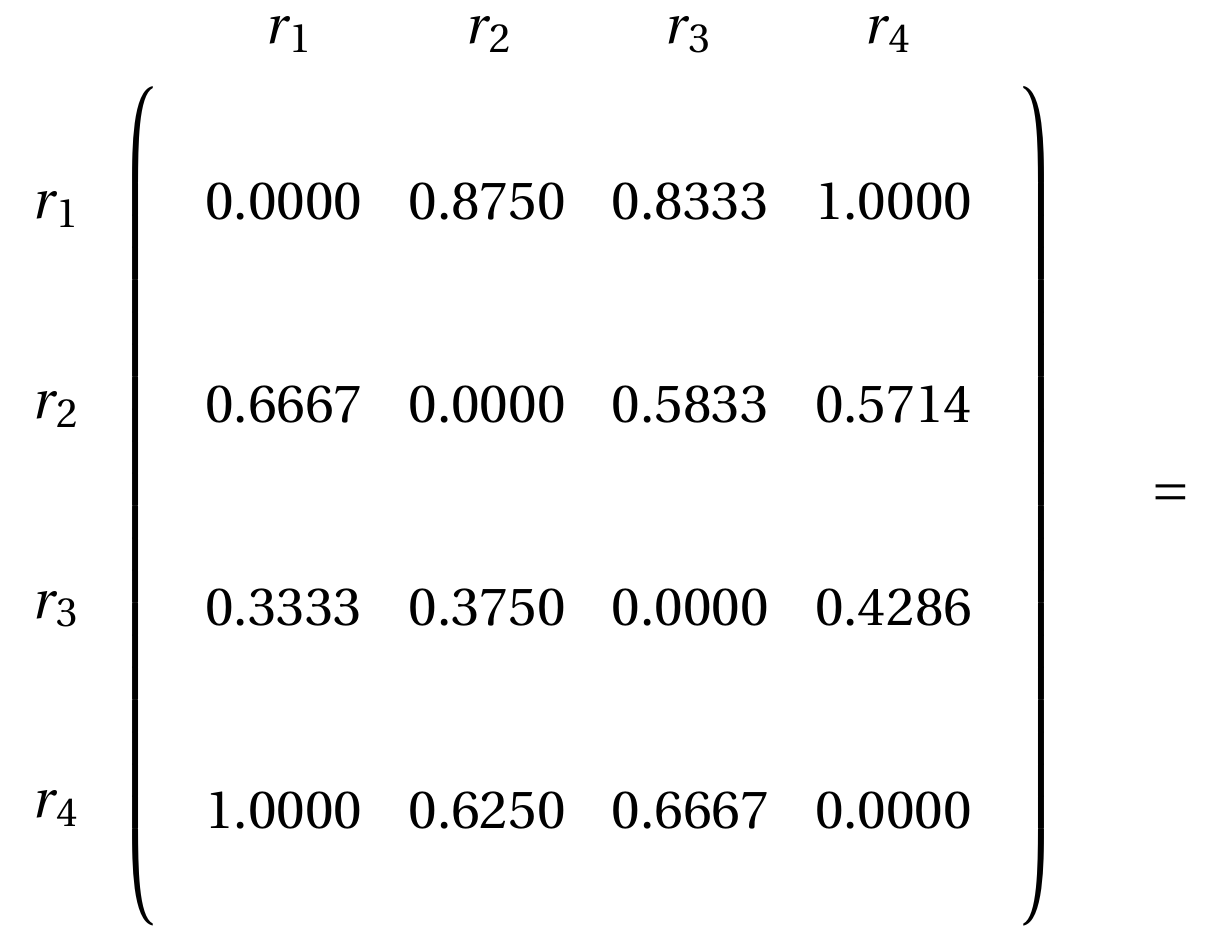}
        }
        \subfigure[$\bm{\eta}_{4 \times 4}$]{%
            \label{fig_ch_dars_eta_eta}
            \includegraphics[scale=0.3]{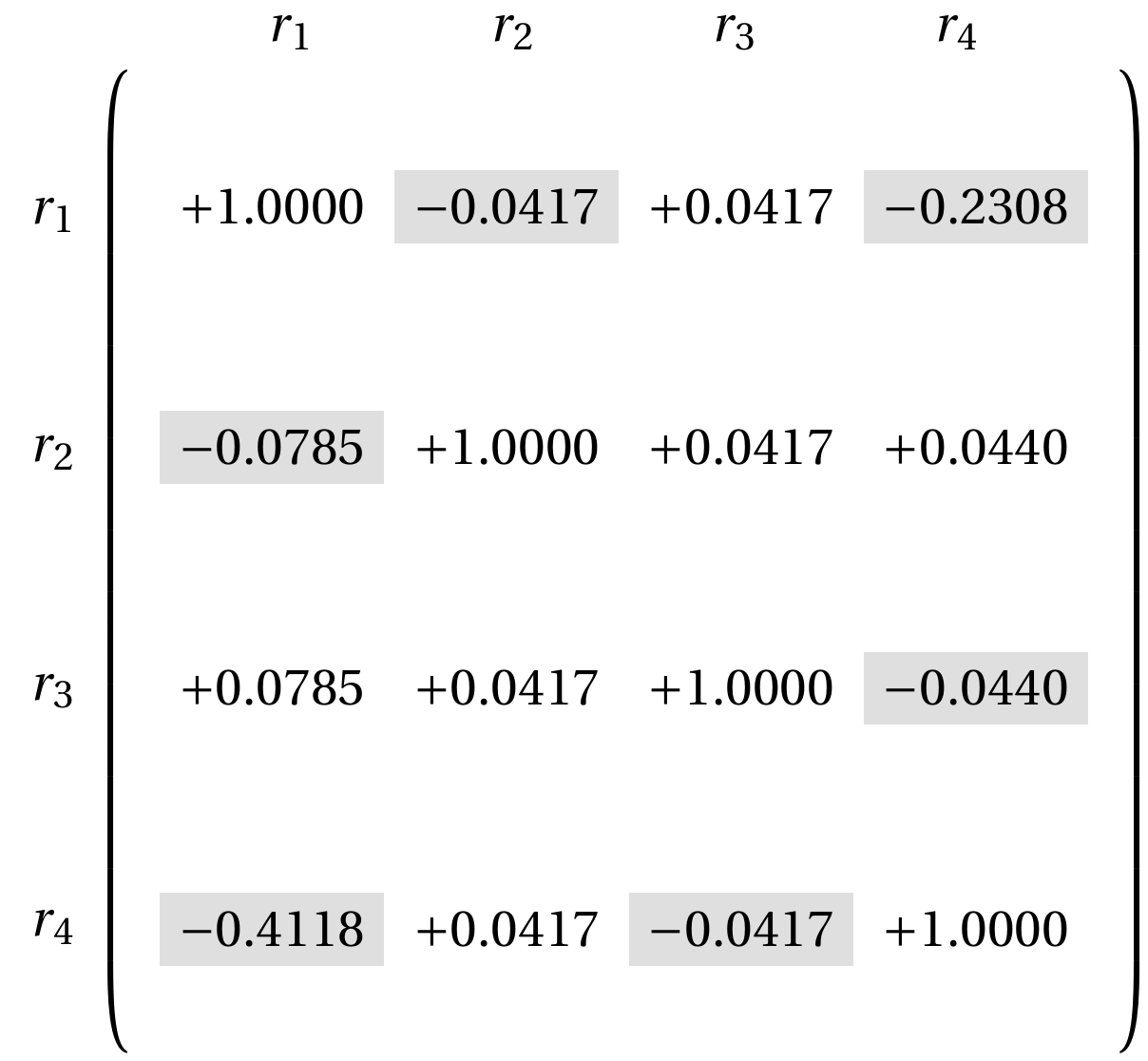}
        }
    \end{center}
    \vspace{-0.25cm}
    \caption{%
        Computing Eells measure for the preference matrix of Figure~\ref{fig_ch_dars_pm}.
    }%
    \label{fig_ch_dars_eta}
\end{figure}

Matrix $P_{4\times4}$ (Figure~\ref{fig_ch_dars_eta_p}) and Matrix $\bar{P}_{4\times4}$ (Figure~\ref{fig_ch_dars_eta_p_bar}) show the strengths of positive and negative causal relations among user preferences for requirements in the preference matrix $M_{4\times 8}$ (Figure~\ref{fig_ch_dars_pm}). For a pair of requirements $r_i$ and $r_j$ with $i \neq j$, an off-diagonal element $p_{i,j}$ ($\bar{p}_{i,j}$) of matrix $P_{4\times4}$ ($\bar{P}_{4\times4}$) denotes the strength of a positive (negative) causal relation from $r_i$ to $r_j$. 

For diagonal elements of $P_{4\times4}$ ($\bar{P}_{4\times4}$) on the other hand, we have $p_{i,i}=p(r_i|r_i)=1$ ($\bar{p}_{i,i}=p(r_i|\bar{r}_i)=0$). Hence, subtracting each element $\bar{p}_{i,j}$ from its corresponding element $p_{i,j}$, where $i \neq j$, gives Eells causal strength $\eta_{i,j}$ for the value dependency from $r_i$ to $r_j$. Diagonal elements, however, may be ignored or set to zero as self-causation is not meaningful here.

\begin{algorithm}[!htb]
    \normalsize
    \caption{Computing Eells measure of strength.}
    \label{ch_dars_alg_identification}
    \begin{algorithmic}[1]
        \REQUIRE \textit{Matrix of user preferences: $M_{n\times k}$} 
        \ENSURE  \textit{Matrix of Eells measure: $\bm{\eta}_{n \times n}$}
        \STATE $P_{n \times n} \leftarrow 0$ 
        \STATE $\bar{P}_{n \times n} \leftarrow 0$ 
        \STATE $\bm{\eta}_{n \times n} \leftarrow 0$ 
        \STATE $\bm{\lambda}{n \times 2n} \leftarrow 0$         
        \FOR{\textbf{each} $r_{j} \in R$}
        \FOR{\textbf{each} $r_{i} \in R$}
        \FOR{\textbf{each} $u_{t} \in U$}
        \IF{$m_{j,t}=1$}
        \IF{$m_{i,t}=1$}
        \STATE $\lambda_{i,j} \leftarrow (\lambda_{i,j} + 1)$ 
        \ELSE
        \STATE $\lambda_{i,j+n} \leftarrow (\lambda_{i,j+n} + 1)$ 
        \ENDIF
        \ENDIF
        \ENDFOR
        \STATE $p_{i,j} \leftarrow (\frac{\lambda_{i,j}}{\lambda_{j,j}})$
        \STATE $\bar{p}_{i,j} \leftarrow (\frac{\lambda_{i,j+n}}{\lambda_{j+n,j+n}})$
        \STATE $\eta_{i,j} \leftarrow  (p_{i,j}-\bar{p}_{i,j})$ 
        \ENDFOR
        \ENDFOR    
    \end{algorithmic}
\end{algorithm}

Algorithm~\ref{ch_dars_alg_identification} specifies the steps for computing the measure of causal strength for a given preference matrix $M_{n\times k}$. In this algorithm, an element $\lambda_{i,j}$ in matrix $\lambda_{n\times 2n}$ counts the number of times that a pair of requirements ($r_i$,$r_j$) are selected together by the users. An element $\lambda_{i,j+n}$ on the other hand, gives the number of times users have selected $r_i$ while ignoring $r_j$. It is clear that, $\lambda_{i,i}$ gives the number of occurrences of $r_i$ in $M_{n\times k}$ while $\lambda_{i,i+n}=0$. 

Given a dataset of $n$ requirements and $t$ user preferences, lines 8 to 14 of Algorithm~\ref{ch_dars_alg_identification} will be executed for each pair of requirements and all gathered user preferences: $O(t \times n^2)$. Moreover, lines 16 to 18 need to be executed for all pairs of requirements. The computational complexity of the algorithm is therefore of $O(n^2)$. The overall complexity of the algorithm, therefore, is of $O(t \times n^2)$.  

\subsection{Testing the Significance of Causal Relations}
\label{ch_dars_identification_testing}

Using measures of interestingness~\citep{geng2006interestingness} is sometimes not sufficient to understand the significance of the relations found among the items of a dataset as explained in~\citep{li2016observational}. In this regard, we have employed the widely adopted measure of association referred to as the \textit{Odds Ratio} to test if causal relations identified based on the Eells measure are significant or not. For a positive (negative) causal relation from requirement $r_j$ to $r_i$, which means the presence of $r_j$ positively (negatively) influences the value of $r_i$, (\ref{Eq_ch_dars_or}) computes the Odds ratio denoted by $\omega(r_i,r_j)$ in which the order of $r_i$ and $r_j$ does not make any difference. Also, $p(r_i,r_j)$ denotes the joint probability of $r_i$ and $r_j$. Similarly, $p(r_i,\bar{r_j})$ gives the joint probability that $r_i$ is selected and $r_j$ is not. 


\begin{align}
\label{Eq_ch_dars_or}
& \omega(r_i,r_j)= \frac{p(r_i,r_j) p(\bar{r_i} ,\bar{r_j})}{p(r_i ,\bar{r_j}) p(\bar{r_i},r_j)} ,\phantom{s}\omega(r_i,r_j) \in (0,\infty)
\end{align}

To test the significance of a causal relation from a requirement $r_j$ to $r_i$, we use the technique used in~\citep{li2016observational} by computing the lower bound ($\omega_{-}$) and the upper bound ($\omega_{+}$) of the confidence interval of the Odds Ratio as given by (\ref{Eq_ch_dars_or_l})-(\ref{Eq_ch_dars_or_h}). In these equations $z^\prime$ is the critical value corresponding to a desired level of confidence. Also, $u$ denotes the total number of user preferences. When we find a lower bound $ \omega_{-} \leq 1$ AND an upper bound $\omega_{+} \geq 1$ for the Odds ratio imply the absence of any significant causal relation from $r_j$ and $r_i$. To exclude insignificant relations, the strengths of those relations will be set to zero.

\begin{align}
\label{Eq_ch_dars_or_l}
\nonumber
\omega_{-} (r_i,r_j)=\\ 
ln \big(\omega (r_i&,r_j) \big) - \frac{z^{\prime}}{\sqrt{u}} \sqrt{\frac{1}{p(r_i , r_j)} + \frac{1}{p(\bar{r_i} , \bar{r_j})} + \frac{1}{p(\bar{r_i} , r_j)} +\frac{1}{p(r_i , \bar{r_j})}} \\
\label{Eq_ch_dars_or_h}
\nonumber
\omega_{+} (r_i,r_j)=\\ 
ln \big(\omega (r_i&,r_j) \big) + \frac{z^{\prime}}{\sqrt{u}} \sqrt{\frac{1}{p(r_i , r_j)} + \frac{1}{p(\bar{r_i} , \bar{r_j})} + \frac{1}{p(\bar{r_i} , r_j)} +\frac{1}{p(r_i , \bar{r_j})}} 
\end{align}

\subsection{Computing the Strengths and Qualities of value Dependencies}
\label{ch_dars_identification_computing}

\begin{figure}[!htb]
    \begin{center}
        \subfigure[$$]{%
            \label{fig_ch_dars_membership_1}
            \includegraphics[scale=0.85]{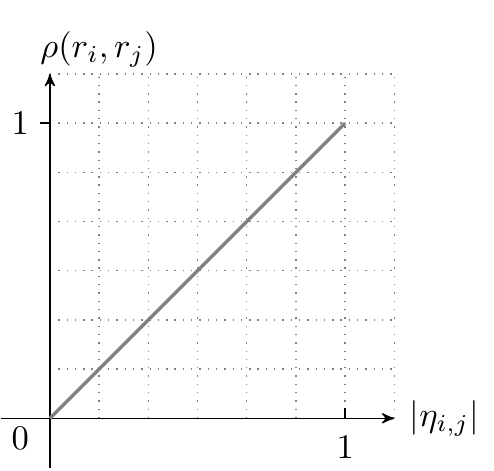}
        }
        \subfigure[$$]{%
            \label{fig_ch_dars_membership_2}
            \includegraphics[scale=0.85]{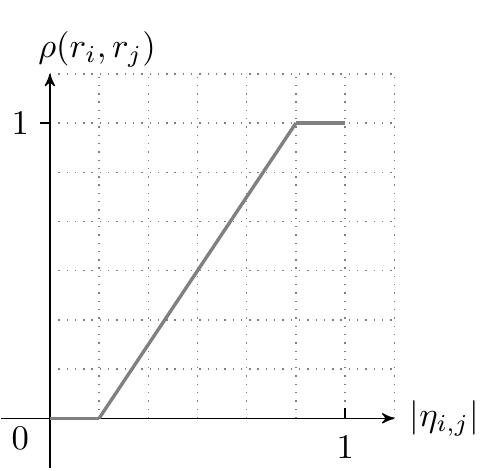}
        }
    \end{center}
    \captionsetup{margin=0ex}
    \caption{%
        Sample membership functions for strengths of value dependencies.
    }%
    \label{fig_ch_dars_membership}
\end{figure}

The strength of an explicit value dependency from a requirement $r_i$ to $r_j$ is computed by (\ref{Eq_ch_dars_strengthMeasure}), which gives a mapping from Eells measure of causal strength $\eta_{i,j}$ to the fuzzy membership function $\rho: R\times R\rightarrow [0,1]$ as given in Figure~\ref{fig_ch_dars_membership}. Significant causal relations which pass the test in Section~\ref{ch_dars_identification_testing} will be considered. 

The fuzzy membership functions, however, may be adjusted to account for the imprecision of value dependencies and suit a particular needs of decision makers. For instance, the membership function of Figure~\ref{fig_ch_dars_membership_1} may be used to ignore ``too weak'' value dependencies while ``too strong'' dependencies are considered as full strength relations, $\rho(r_i,r_j)=1$. Different membership functions and measures of causal strength may be used by decision makers resulting in a set of optimal solutions to choose from. 

\begin{align}
\label{Eq_ch_dars_strengthMeasure}
& \rho(r_i,r_j)= |\eta_{i,j}|
\end{align}

\begin{align}
\label{Eq_ch_dars_qualityMeasure}
& \sigma(r_i,r_j) =  \begin{cases}
+ & \text{if }\phantom{s}  \eta_{i,j} > 0 \\
- & \text{if }\phantom{s}  \eta_{i,j} < 0 \\
\pm & \text{if }\phantom{s} \eta_{i,j} = 0 \\
\end{cases}
\end{align}

As given by (\ref{Eq_ch_dars_qualityMeasure}), $\eta_{i,j}>0$ indicates that the strength of the positive causal relation from $r_i$ to $r_j$ is greater than the strength of its corresponding negative causal relation: $p(r_i|r_j) > p(r_i|\neg r_j)$ and therefore, the quality of $(r_i,r_j)$ is positive ($\sigma(r_i,r_j)=+$). Similarly, $\eta_{i,j}<0$ indicates $p(r_i|\neg r_j) > p(r_i|r_j) \rightarrow \sigma(r_i,r_j)=-$. Also, $p(r_i|r_j) - p(r_i|\neg r_j)=0$ specifies that the quality of the zero-strength value dependency $(r_i,r_j)$ is non-specified ($\sigma(r_i,r_j)=\pm$).

\subsection{Value Implications of Precedence Dependencies}
\label{ch_dars_identification_precedence}

As explained earlier, precedence dependencies among requirements such as \textit{requires} and \textit{conflicts-with} and their value implications need to be considered in requirements selection. For instance, a requirement $r_i$ requires (conflicts-with) $r_j$ implies that the value of $r_i$ fully relies on selecting (ignoring) $r_j$. This may not be captured by value dependencies identified from user preferences. 

Hence, it is important to not only consider user preferences in the identification of explicit value dependencies but to take into account the value implications of precedence dependencies and consider them in a requirements selection. This can be achieved by modeling the precedence dependencies using a \textit{Precedence Dependency Graph} (PDG) as introduced in Definition~\ref{def_ch_dars_PDG}.

\begin{mydef}
    \label{def_ch_dars_PDG}
    \textit{The Precedence Dependency Graph} (PDG). A PDG is a signed directed graph $G=(R,W)$ in which $R=\{r_1,...,r_n\}$ denotes the graph nodes (requirements) and $W(r_i,r_j)\in {-1,0,1}$ specifies the presence or absence of a precedence dependency from $r_i$ to $r_j$. $W(r_i,r_j)=1$ ($W(r_i,r_j)=-1$) specifies a positive (negative) precedence dependency from $r_i$ to $r_j$ meaning that $r_i$ \textit{requires} (\textit{conflicts-with}) $r_j$. Finally $W(r_i,r_j)=0$ specifies the absence of any precedence dependency from requirement $r_i$ to $r_j$. 
\end{mydef}

\begin{align}
\label{Eq_ch_dars_pdl}
&PDL(G)=\frac{k}{\Perm{n}{2}}=\frac{k}{n(n-1)} \\
\label{Eq_ch_dars_npdl}
&NPDL(G)=\frac{j}{k}
\end{align}

Hence, precedence dependencies of a software project can be captured by a PDG and mathematically modeled in terms of the precedence constraints of the optimization model used for a requirements selection. It is clear that increasing the precedence dependencies among requirements limits the number of choices and therefore reduce the number of feasible solutions (requirement subsets). To measure the level of precedence dependencies among requirements of a PDG, we have defined the \textit{Precedence Dependency Level} (PDL) and the \textit{Negative Precedence Dependency Level} (NPDL) as given by (\ref{Eq_ch_dars_pdl}) and (\ref{Eq_ch_dars_npdl}) respectively.

The PDL of a precedence dependency graph $G$ with $n$ nodes (requirements)  is computed by dividing the total number of the precedence dependencies ($k$) among the nodes of $G$ by the maximum number of the potential precedence dependencies in $G$ ($n(n-1)$). Also, the NPDL of $G$ is computed by dividing the number of negative precedence dependencies ($j$) by the total number of the positive and negative precedence dependencies.

%% file: modeling.tex
\section{Modeling Value Dependencies by Fuzzy Graphs}
\label{ch_dars_modeling}
\hypertarget{ch_dars_modeling}{ }	

Fuzzy logic and Fuzzy graphs~\cite{Mathew_strong_2013} have been widely adopted in decision making and expert systems~\citep{rosenfeld_fuzzygraph_1975} as they contribute to more accurate models by taking into account the imprecision of real-world problems~\citep{Mathew_strong_2013,mougouei2012goal,mougouei2012measuring,mougouei2012evaluating,mougouei2013goal,mougouei2013fuzzy}. Fuzzy logic has been adopted in requirement selection for capturing the partiality of requirements~\cite{mougouei2019fuzzy,mougouei2015partial}. Also, Fuzzy graphs have, particularly, demonstrated useful in capturing the imprecision of dependency relations in software~\citep{ngo_fuzzy_2005_structural,ngo2005measuring,mougouei2017modeling,mougouei2018mathematical}. Ngo-The \textit{et al.}, exploited fuzzy graphs for modeling dependency satisfaction in release planning~\citep{ngo_fuzzy_2005_structural} and capturing the imprecision of coupling dependencies among requirements~\citep{ngo2005measuring}. 

Moreover, Wang \textit{et al.}~\citep{wang_simulation_2012} adopted linguistic fuzzy terms to capture the variances of strengths of dependencies among software requirements. In this section we discuss modeling value dependencies by fuzzy graphs and identification of implicit value dependencies among requirements. We further use the algebraic structure of fuzzy graphs to compute the influences of requirements on the values of each other.   

\subsection{Value Dependency Graphs}
\label{ch_dars_modeling_vdg}
 
To account for the imprecision of value dependencies, we have introduced \textit{Value Dependency Graphs} (VDGs) based on fuzzy graphs for modeling value dependencies and their characteristics. We have specially modified the classical definition of fuzzy graphs to consider not only the strength but also the quality (positive or negative) of value dependencies as given by Definition~\ref{def_vdg}. 

\begin{mydef}
	\label{def_vdg}
	\textit{The Value Dependency Graph} (VDG) is a signed directed fuzzy graph~\citep{Wasserman1994} $G=(R,\sigma,\rho)$ where, requirements $R:\{r_1,...,r_n\}$ constitutes the graph nodes. Also, the qualitative function $\sigma(r_i,r_j) \rightarrow \{+,-,\pm\}$ and the membership function $\rho: (r_i,r_j)\rightarrow [0,1]$ denote the quality and the strength of the explicit value dependency (edge of the graph) from $r_i$ to $r_j$ receptively. Moreover, $\rho(r_i,r_j)=0$ denotes the absence of any explicit value dependency from $r_i$ to $r_j$. In that case we have $\sigma(r_i,r_j)=\pm$, where $\pm$ denotes the quality of the dependency is non-specified.  
\end{mydef}

\begin{figure}[!htb]
	\begin{center}
		\includegraphics[scale=0.8]{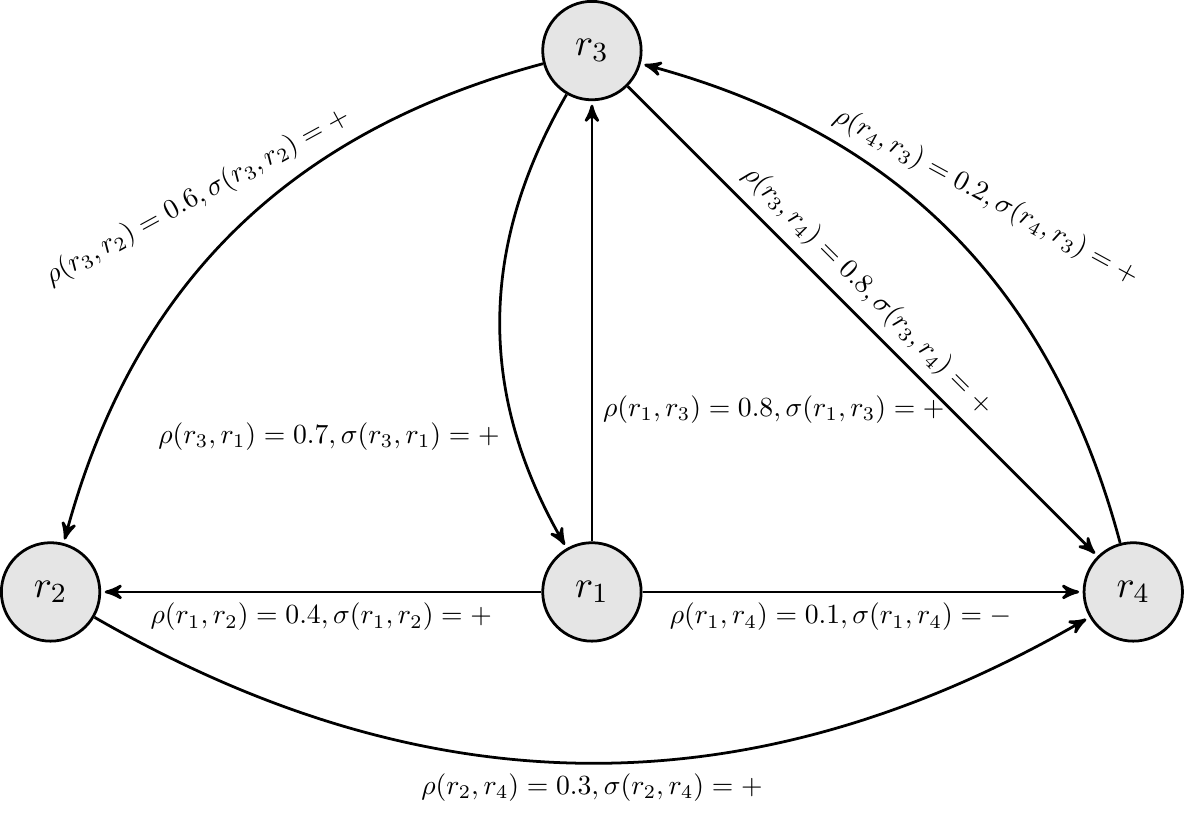}
	\end{center}
	\caption{%
		A sample value dependency graph.}%
	\label{fig_ch_dars_ex_vdg}
\end{figure}

For instance, in the value dependency graph of Figure~\ref{fig_ch_dars_ex_vdg} $\sigma(r_1,r_2)=+$ and $\rho(r_1,r_2)=0.4$ specifies a positive value dependency from $r_1$ to $r_2$ with strength $ 0.4 $. That is selecting $r_2$ has an explicit positive influence on the value of $r_1$. 

\subsection{Value Dependencies in VDGs}
\label{ch_dars_modeling_vdg}

In Section~\ref{ch_dars_identification} we introduced an automated technique for identification of explicit value dependencies and their characteristics (quality and strength) from user preferences. Definition~\ref{def_vdg_valuedepndencies} provides a more comprehensive definition of value dependencies that includes both explicit and implicit value dependencies among the requirements of software based on the algebraic structure of fuzzy graphs. 

\begin{mydef}
	\label{def_vdg_valuedepndencies}
	\textit{Value Dependencies}. 
	A value dependency in a value dependency graph $G=(R,\sigma,\rho)$ is defined as a sequence of requirements $d_i:\big(r(0),...,r(k)\big)$ such that $\forall r(j) \in d_i$, $1 \leq j \leq k$ we have $\rho\big(r(j-1),r(j)\big) \neq 0$. $j\geq 0$ is the sequence of the $j^{th}$ requirement (node) denoted as $r(j)$ on the dependency path. A consecutive pair $\big(r(j-1),r(j)\big)$ specifies an explicit value dependency. 
\end{mydef}

\begin{align}
\label{Eq_ch_dars_vdg_strength}
&\forall d_i:\big(r(0),...,r(k)\big): \rho(d_i) = \bigwedge_{j=1}^{k}\text{ }\rho\big(r(j-1),r(j)\big) \\
\label{Eq_ch_dars_vdg_quality}
&\forall d_i:\big(r(0),...,r(k)\big): \sigma(d_i) = \prod_{j=1}^{k}\text{ }\sigma\big(r(j-1),r(j)\big)
\end{align}

Equation (\ref{Eq_ch_dars_vdg_strength}) computes the strength of a value dependency $d_i:\big(r(0),...,r(k)\big)$ by finding the strength of the weakest of the $k$ explicit dependencies on $d_i$. Fuzzy operator $\wedge$ denotes Zadeh's~\citep{zadeh_fuzzysets_1965} AND operation (infimum). The quality (positive or negative) of a value dependency $d_i:\big(r(0),...,r(k)\big)$ is calculated by qualitative serial inference~\citep{de1984qualitative,wellman1990formulation,kusiak_1995_dependency} as given by (\ref{Eq_ch_dars_vdg_quality}) and Table~\ref{table_ch_dars_inference}. Inferences in Table~\ref{table_ch_dars_inference} are informally proved by Wellman~\citep{wellman1990formulation} and Kleer~\citep{de1984qualitative}. 

\begin{table}[!htb]
	\caption{Qualitative serial inference in VDGs.}
	\label{table_ch_dars_inference}
	\centering
	\input{table_inference}

\end{table}


Let $D=\{d_1,d_2,..., d_m\}$ be the set of all value dependencies from $r_i \in R$ to $r_j \in R$ in a VDG $G=(R,\sigma,\rho)$, where positive and negative dependencies can simultaneously exist from $r_i$ to $r_j$. The strength of all positive value dependencies from $r_i$ to $r_j$, denoted by $\rho^{+\infty}(r_i,r_j)$, is calculated by (\ref{Eq_ch_dars_ultimate_strength_positive}), that is to find the strength of the strongest positive dependency~\citep{rosenfeld_fuzzygraph_1975} from $r_i$ to $r_j$. Fuzzy operators $\wedge$ and $\vee$ denote Zadeh's~\citep{zadeh_fuzzysets_1965} fuzzy AND and OR operations respectively. Analogously, the strength of all negative value dependencies from $r_i$ to $r_j$ is denoted by $\rho^{-\infty}(r_i,r_j)$ and calculated by (\ref{Eq_ch_dars_ultimate_strength_negative}).

\begin{align}
\label{Eq_ch_dars_ultimate_strength_positive}
&\rho^{+\infty}(r_i,r_j) = \bigvee_{d_m\in D, \sigma(d_m)=+} \text{ } \rho(d_m) \\[2pt]
\label{Eq_ch_dars_ultimate_strength_negative}
&\rho^{-\infty}(r_i,r_j) = \bigvee_{d_m\in D, \sigma(d_i)=-} \text{ } \rho(d_m) 
\end{align}

A brute-force approach to computing $\rho^{+\infty}(r_i,r_j)$ or $\rho^{-\infty}(r_i,r_j)$ needs to calculate the strengths of all paths from $r_i$ to $r_j$, which is of complexity of $O(n!)$ for $n$ requirements (VDG nodes). To avoid such complexity, we devised a modified version of Floyd-Warshall~\citep{floyd_1962} algorithm (Algorithm~\ref{alg_ch_dars_strength}) that computes $\rho^{+\infty}(r_i,r_j)$ and $\rho^{-\infty}(r_i,r_j)$ for all pairs of requirements $(r_i,r_j),\text{ }r_i,r_j \in R:\{r_1,...,r_n\}$ in polynomial time: $O(n^3)$. For each pair of requirements $(r_i,r_j)$ in a VDG $G=(R,\sigma,\rho)$, lines $18$ to $35$ of Algorithm~\ref{alg_ch_dars_strength} find the strength of all positive (negative) value dependencies from $r_i$ to $r_j$.

\begin{algorithm}
	\footnotesize
	\caption{Calculating the strengths of value dependencies.}
	\label{alg_ch_dars_strength}
	\begin{algorithmic}[1]
		\REQUIRE VDG $G=(R,\sigma,\rho)$
		\ENSURE $\rho^{+\infty}, \rho^{-\infty}$
		\FOR{\textbf{each} $r_i \in R$}
		\FOR{\textbf{each} $r_j \in R$}
		\STATE $\rho^{+\infty}(r_i,r_j) \leftarrow \rho^{-\infty}(r_i,r_j) \leftarrow -\infty$ 
		\ENDFOR
		\ENDFOR
		\FOR{\textbf{each} $r_i \in R$}
		\STATE $\rho(r_i,r_i)^{+\infty} \leftarrow \rho(r_i,r_i)^{-\infty} \leftarrow 0$
		\ENDFOR
		\FOR{\textbf{each} $r_i \in R$}
		\FOR{\textbf{each} $r_j \in R$}
		\IF{$\sigma(r_i,r_j) = +$}
		\STATE $\rho^{+\infty}(r_i,r_j) \leftarrow \rho(r_i,r_j)$
		\ELSIF{$\sigma(r_i,r_j) = -$}
		\STATE $\rho^{-\infty}(r_i,r_j) \leftarrow \rho(r_i,r_j)$
		\ENDIF
		\ENDFOR
		\ENDFOR
		\FOR{\textbf{each} $r_k \in R$}
		\FOR{\textbf{each} $r_i \in R$}
		\FOR{\textbf{each} $r_j \in R$}
		\IF{$min\big(\rho^{+\infty}(r_i,r_k), \rho^{+\infty}(r_k,r_j)\big) > \rho^{+\infty}(r_i,r_j)$}
		\STATE $\rho^{+\infty}(r_i,r_j) \leftarrow  min(\rho^{+\infty}(r_i,r_k), \rho^{+\infty}(r_k,r_j))$
		\ENDIF
		\IF{$min\big(\rho^{-\infty}(r_i,r_k), \rho^{-\infty}(r_k,r_j)\big) > \rho^{+\infty}(r_i,r_j)$}
		\STATE $\rho^{+\infty}(r_i,r_j) \leftarrow  min(\rho^{-\infty}(r_i,r_k), \rho^{-\infty}(r_k,r_j))$
		\ENDIF
		\IF{$min\big(\rho^{+\infty}(r_i,r_k), \rho^{-\infty}(r_k,r_j)\big) > \rho^{-\infty}(r_i,r_j)$}
		\STATE $\rho^{-\infty}(r_i,r_j) \leftarrow  min(\rho^{+\infty}(r_i,r_k), \rho^{-\infty}(r_k,r_j))$
		\ENDIF
		\IF{$min\big(\rho^{-\infty}(r_i,r_k), \rho^{+\infty}(r_k,r_j)\big) > \rho^{-\infty}(r_i,r_j)$}
		\STATE $\rho^{-\infty}(r_i,r_j) \leftarrow  min(\rho^{-\infty}(r_i,r_k), \rho^{+\infty}(r_k,r_j))$
		\ENDIF
		\ENDFOR
		\ENDFOR
		\ENDFOR
	\end{algorithmic}
\end{algorithm}

\begin{align}
\label{Eq_ch_dars_influence}
I_{i,j} = \rho^{+\infty}(r_i,r_j)-\rho^{-\infty}(r_i,r_j) 
\end{align}

The overall strength of all positive and negative value dependencies from $r_i$ to $r_j$ is referred to as the \textit{Influence} of $r_j$ on the value of $r_i$ and denoted by $I_{i,j}$. $I_{i,j}$ as given by (\ref{Eq_ch_dars_influence}) is calculated by subtracting the strength of all negative value dependencies from $r_i$ to $r_j$ ($\rho^{-\infty}(r_i,r_j)$) from the strength of all positive value dependencies from $r_i$ to $r_j$ ($\rho^{+\infty}(r_i,r_j)$). It is clear that $I_{i,j}\in[-1,1]$. $I_{i,j}>0$ states that $r_j$ positively influences the value of $r_i$ whereas $I_{i,j}<0$ indicates a negative influence from $r_j$ on $r_i$.

%
%

\begin{exmp}
	\label{ex_ultimate_strength}
	Let $D=\{d_1:(r_1,r_2,r_4),d_2:(r_1,r_3,r_4),d_3:(r_1,r_4)\}$ specify value dependencies from requirement $r_1$ to $r_4$ in Figure \ref{fig_ch_dars_ex_vdg}. Using (\ref{Eq_ch_dars_vdg_quality}), qualities of $d_1$ to $d_3$ are computed as: $\sigma(d_1)=\Pi(+,+)=+$, $\sigma(d_2)=\Pi(+,+)=+$, and $\sigma(d_3)=\Pi(-)=-$. Strengths are calculated by (\ref{Eq_ch_dars_vdg_strength}) as: $\rho(d_1)=\wedge\big(\rho(r_1,r_2),\rho(r_2,r_4)\big)=min(0.4,0.3)$, $\rho(d_2)=\wedge\big(\rho(r_1,r_3),\rho(r_3,r_4)\big)=min(0.8,0.8)$, $\rho(d_3)=min(0.1)$. Using~(\ref{Eq_ch_dars_ultimate_strength_positive})and~(\ref{Eq_ch_dars_ultimate_strength_negative}) then we have $\rho(r_1,r_4)^{+\infty} = \vee(\rho(d_1),\rho(d_2))=max(0.3,0.8)$ and $\rho^{-\infty}(r_1,r_4) = max(\rho(d_3))$. Therefore, we have $I_{1,4} = \rho(r_1,r_4)^{+\infty}-\rho(r_1,r_4)^{-\infty}=0.7$ which means the positive influence of $r_4$ on the value of $r_1$ prevails. Table~\ref{table_ch_dars_ex_overall} lists influences of requirements in the VDG of Figure~\ref{fig_ch_dars_ex_vdg} on the value of each other.
\end{exmp}
\begin{table}[!htb]
	\centering
	\caption{Overall influences computed for VDG of Figure~\ref{fig_ch_dars_ex_vdg}.}
	\label{table_ch_dars_ex_overall}
	\input{table_ex_strengths}

\end{table}

\begin{mydef}
	\label{def_frig_VDL}
	\textit{Value Dependency Level (VDL) and Negative Value Dependency Level (NVDL)}. Let $G=(R,\sigma,\rho)$ be a VDG with $R=\{r_1,...,r_n\}$, $k$ be the total number of explicit value dependencies in $G$, and $m$ be the total number of negative explicit value dependencies. Then the VDL and NVDL of $G$ are derived by (\ref{Eq_ch_dars_vdl}) and (\ref{Eq_ch_dars_nvdl}) respectively. 
\end{mydef}

\begin{align}
\label{Eq_ch_dars_vdl}
&VDL(G)=\frac{k}{\Perm{n}{2}}=\frac{k}{n (n-1)} \\
\label{Eq_ch_dars_nvdl}
&NVDL(G)=\frac{m}{k}
\end{align}

\begin{exmp}
	\label{ex_vdl}
	For the value dependency graph $G$ of Figure~\ref{fig_ch_dars_ex_vdg} we have $n=4$, $k=8$, and $m=1$. $VDL(G)$ is derived by~(\ref{Eq_ch_dars_vdl}) as: $VDL(G)= \frac{8}{4\times 3} = \frac{8}{12} \approxeq 0.67$. Also we have from Equation~(\ref{Eq_ch_dars_nvdl}), $NVDL(G)=\frac{1}{8}=0.125$.
\end{exmp}

%% file: table_inference.tex
\resizebox {0.55\textwidth }{!}{
	\begin{tabular}{cc|ccc}
		\toprule[1.5pt]
		\multicolumn{2}{r|}{\multirow{2}[1]{*}{ $\sigma\big(r(j-1),r(j),r(j+1)\big)$}} &
		\multicolumn{3}{c}{$\sigma\big(r(j),r(j+1)\big)$}
		\\
		\multicolumn{2}{r|}{} &
		$+$ &
		$-$ &
		$\pm$
		\bigstrut[b]\\
		\hline
		\multicolumn{1}{c}{\multirow{3}[1]{*}{$\sigma\big(r(j-1),r(j)\big)$}} &
		$+$ &
		$+$ &
		$-$ &
		$\pm$
		\bigstrut[t]\\
		\multicolumn{1}{c}{} &
		$-$ &
		$-$ &
		$+$ &
		$\pm$
		\\
		\multicolumn{1}{c}{} &
		$\pm$ &
		$\pm$ &
		$\pm$ &
		$\pm$
		\\
    \bottomrule[1.5pt]
	\end{tabular}%
	}

%% file: table_ex_strengths.tex
\resizebox {1\textwidth }{!}{
	\begin{tabular}{lcccc}
		\toprule[1.5pt]
		\textbf{\cellcolor{white}\textcolor{black}{$I_{i,j}=\rho(r_i,r_j)^{+\infty}-\rho(r_i,r_j)^{-\infty}$}}&
		\textbf{\cellcolor{white}\textcolor{black}{$r_1$}}&
		\textbf{\cellcolor{white}\textcolor{black}{$r_2$}}&
		\textbf{\cellcolor{white}\textcolor{black}{$r_3$}}&
		\textbf{\cellcolor{white}\textcolor{black}{$r_4$}}
		\\ \midrule
		\textbf{$r_1$}\unboldmath{} &
		$0.0-0.0=0.0$ &
		$0.6-0.1=0.5$ &
		$0.8-0.1=0.7$ &
		$0.8-0.1=0.7$
		\\
		\textbf{$r_2$}\unboldmath{} &
		$0.2-0.0=0.2$ &
		$0.0-0.0=0.0$ &
		$0.2-0.0=0.2$ &
		$0.3-0.0=0.3$
		\\
		\textbf{$r_3$}\unboldmath{} &
		$0.7-0.1=0.6$ &
		$0.6-0.1=0.5$ &
		$0.0-0.0=0.0$ &
		$0.8-0.1=0.7$
		\\
		\textbf{$r_4$}\unboldmath{} &
		$0.2-0.0=0.2$ &
		$0.2-0.0=0.2$ &
		$0.2-0.0=0.2$ &
		$0.0-0.0=0.0$
		\\ \bottomrule[1.5pt]
	\end{tabular}
}

%% file: selection.tex
\section{Integrating Value Dependencies into Requirements Selection}
\label{ch_dars_selection}

%% file: selection_ov.tex
\subsection{Overall Value of a Subset of Requirements}
\label{ch_dars_selection_ov}

This section details our proposed measure for the economic worth of a selected subset of requirements (software product) i.e. overall value (OV) as an alternative to the accumulated value (AV) and the expected value (EV) of that subset. The formulation of overall value in this section takes into account user preferences for selected requirements as well as the impacts of value dependencies on the values of requirements. 

Value dependencies as explained in Section~\ref{ch_dars_identification} are identified based on causal relations among user preferences. Section~\ref{ch_dars_identification} presented an automated technique for identification of value dependencies among requirements. Then, algorithm~\ref{alg_ch_dars_strength} was used to infer implicit value dependencies and compute the influences of requirements on the values of each other based on the algebraic structure of fuzzy graphs.  

To compute the overall values of selected requirements, (\ref{eq_ch_dars_penalty})-(\ref{eq_ch_dars_penalty_c1}) give the penalty of ignoring (selecting) requirements with positive (negative) influence on the values of selected requirements. $\theta_{i}$ in this equation denotes the penalty for a requirement $r_i$, $n$ denotes the number of requirements, and $x_j$ specifies whether a requirement $r_j$ is selected ($x_j=1$) or not ($x_j=0$). Also, $I_{i,j}$, as in (\ref{Eq_ch_dars_influence}), gives the positive or negative influence of $r_j$ on the value of $r_i$.  

\begin{align}
\label{eq_ch_dars_penalty}
\nonumber
\theta_{i}= &\displaystyle \bigvee_{j=1}^{n} \bigg(\frac{x_j\big(\lvert I_{i,j} \rvert-I_{i,j}\big) + (1-x_j)\big(\lvert I_{i,j}\rvert+I_{i,j}\big)}{2}\bigg)&&=\\ 
&\displaystyle \bigvee_{j=1}^{n} \bigg(\frac{\lvert I_{i,j} \rvert + (1-2x_j)I_{i,j}}{2}\bigg),&& i \neq j =1,...,n \\
\label{eq_ch_dars_penalty_c1}
& x_j \in\{0,1\},\quad \quad \quad \quad \quad \quad \quad  \hspace{0.2em} && j=1,...,n 
\end{align}

We made use of the algebraic structure of fuzzy graphs for computing the influences of requirements on the values of each other as explained in Section~\ref{ch_dars_modeling}. Accordingly, $\theta_i$ is computed using the fuzzy OR operator which is to take supremum over the strengths of all ignored positive dependencies and selected negative dependencies of $r_i$ in its corresponding value dependency graph. Overall values of selected requirements thus can be computed by (\ref{eq_ch_dars_vprime}), where $v_i^{\prime}$ denotes the overall value of a requirement $r_i$, $E(v_i)$ specifies the expected value of $r_i$, and $\theta_{i}$ denotes the penalty of ignoring (selecting) positive (negative) value dependencies of $r_i$. 

Equation (\ref{eq_ch_dars_ov}) derives the overall value of a software product with $n$ requirements, where cost and expected value of a requirements $r_i$ are denoted by $c_i$ and $E(v_i)$ respectively. Decision variable $x_i$ specifies whether $r_i$ is selected ($x_i=1$) or not ($x_i=0$). $E(V_i)$ is computed by (\ref{eq_ch_dars_expected}), where $v_i$ denotes the estimated (nominal) value of $r_i$. Also $p(r_i)$/$p(\bar{r_i})$ specify the probability that users select/ignore a requirement $r_i$. 

\vspace{-0.5cm}
\begin{align}
\label{eq_ch_dars_expected}
&E(v_i) = p(r_i)\times v_i + p(\bar{r_i})\times 0= p(r_i)\times v_i
\end{align}

For a requirement $r_i$, $\theta_i$ specifies the penalty of ignoring (selecting) requirements with positive (negative) influence on the expected value of $r_i$ as explained earlier. $\theta_iv_i$ in (\ref{eq_ch_dars_ov}) therefore, gives the value loss for a requirement $r_i$ as a result of ignoring (selecting) requirements that positively (negatively) impact user preferences for $r_i$ and consequently its expected value.

\begin{align}
\label{eq_ch_dars_vprime}
& v_i^{\prime} = (1-\theta_i)E(v_i)
\end{align}
\begin{align}
\label{eq_ch_dars_ov}
&OV = \sum_{i=1}^{n} x_i (1-\theta_i)E(v_i), \textit{ } x_i \in \{0,1\}
\end{align}

\begin{exmp}
	\label{ex_ch_dars_overall}
	Consider finding penalties for requirements of Figure \ref{fig_ch_dars_ex_vdg}, where $r_4$ is not selected ($x_1=x_2=x_3=1,x_4=0$). From Table~\ref{table_ch_dars_ex_overall} we have $I_{1,4}=I_{3,4}=0.7,I_{2,4}=0.3,I_{4,4}=0.0$. As such, based on~(\ref{eq_ch_dars_penalty}) penalties are computed: $\theta_{1}= \vee(\frac{\lvert 0.0 \rvert +(1-2(1))(0.0)}{2}$, $\frac{\lvert 0.45 \rvert +(1-2(1))(0.5)}{2}$, $\frac{\lvert 0.7 \rvert +(1-2(1))(0.7)}{2}$, $\frac{\lvert 0.7 \rvert +(1-2(0))(0.7)}{2}) =0.7$. Similarly, we have $\theta_2=0.3, \theta_3=0.7$. Therefore, the overall value of the selected requirements $r_1,r_2,r_3$ is derived by~(\ref{eq_ch_dars_ov}) as: $ OV (s_1) = (1-0.7)E(v_1) + (1-0.3)E(v_2) + (1-0.7)E(v_3)$. 
\end{exmp}

%% file: selection_ilp.tex
\subsection{The Integer Linear Programming Model}
\label{ch_dars_selection_ilp}

This section presents our proposed integer linear programming (ILP) model for optimizing the overall value of a software product. The overall value of a requirement subset, as given by (\ref{eq_ch_dars_ov}), considers user preferences and the impacts of value dependencies on the expected values of the selected requirements. The proposed ILP model hence embeds user preferences and value dependencies into requirements selection by optimizing the overall value of a software product.  

Equations (\ref{Eq_ch_dars_dars})-(\ref{Eq_ch_dars_dars_c5}) give our proposed integer programming model as a main component of DARS. In these equations, $x_i$ is a selection variable denoting whether a requirement $r_i$ is selected ($x_i=1$) or ignored ($x_i=0$). Also $\theta_i$ in (\ref{eq_ch_dars_penalty}) specifies the penalty of a requirement $r_i$, which is the extent to which the expected value of $r_i$ is impacted by ignoring (selecting) requirements with positive (negative) influences on the value of $r_i$. Constraint~(\ref{Eq_ch_dars_dars_c2}) on the other hand accounts for precedence dependencies among requirements and the value implications of those dependencies. 

\begin{align}
\label{Eq_ch_dars_dars}
& \text{Maximize } \sum_{i=1}^{n} x_i (1-\theta_i) E(v_i)\\[5pt]
\label{Eq_ch_dars_dars_c1}
&\text{Subject to} \sum_{i=1}^{n} c_i x_i \leq b\\[5pt] 
\label{Eq_ch_dars_dars_c2}
& \begin{cases}
x_i \le x_j  & r_j \text{ precedes } r_i \\[5pt]
x_i \le 1-x_j& r_i \text{ conflicts with } r_j,\text{ }i\neq j= 1,...,n\\[5pt]
\end{cases}\\[5pt]
\label{Eq_ch_dars_dars_c3}
& \theta_{i} \geq \bigg(\frac{\lvert I_{i,j} \rvert + (1-2x_j)I_{i,j}}{2}\bigg),& i\neq j = 1,...,n\\[5pt]
\label{Eq_ch_dars_dars_c4}
&\text{ }x_i \in \{0,1\},& i = 1,...,n \\[5pt]
\label{Eq_ch_dars_dars_c5}
&\text{ } 0 \leq \theta_i \leq 1,& i = 1,...,n
\end{align}

Moreover, for a requirement $r_i$, $\theta_i$ depends on the selection variable $x_j$ and the strength of positive (negative) value dependencies as given by~(\ref{eq_ch_dars_penalty}).

Since $I_{i,j}$ is computed by (\ref{Eq_ch_dars_influence}) we can restate $\theta_i$ as a function of $x_j$: $\theta_i=f(x_j)$. The objective function~(\ref{Eq_ch_dars_dars}), thus, can be restated as $\text{Maximize } \sum_{i=1}^{n} x_i E(v_i) - x_if(x_j)E(v_i)$ where $x_if(x_j)E(v_i)$ is a quadratic non-linear expression~\citep{boyd2004convex}. Equations (\ref{Eq_ch_dars_dars})-(\ref{Eq_ch_dars_dars_c3}), on the other hand, denote a convex optimization problem as the model maximizes a concave objective function with linear constraints. 

\begin{align}
\label{Eq_ch_dars_dars_linear}
&\text{Maximize }  \sum_{i=1}^{n} x_i E(v_i) - y_i E(v_i)\\[5pt]
\label{Eq_ch_dars_dars_linear_c1}
&\text{Subject to} \sum_{i=1}^{n} c_i x_i \leq b\\[5pt]
\label{Eq_ch_dars_dars_linear_c2}
& \begin{cases}
x_i \le x_j  & r_j \text{ precedes } r_i \\[5pt]
x_i \le 1-x_j& r_i \text{ conflicts with } r_j,\text{ }i\neq j= 1,...,n
\end{cases}\\[5pt]
\label{Eq_ch_dars_dars_linear_c3}
& \theta_{i}\geq \bigg(\frac{\lvert I_{i,j} \rvert + (1-2x_j)I_{i,j}}{2}\bigg),& i\neq j = 1,...,n\\
\label{Eq_ch_dars_dars_linear_c4}
& -g_i \leq x_i \leq  g_i,& i=1,...,n\\[5pt]
\label{Eq_ch_dars_dars_linear_c5}
& 1-(1-g_i) \leq x_i \leq 1+(1-g_i),& i=1,...,n\\[5pt]
\label{Eq_ch_dars_dars_linear_c6}
& -g_i \leq y_i \leq g_i,& i=1,...,n\\
\label{Eq_ch_dars_dars_linear_c7}
& -(1-g_i)\leq(y_i-\theta_i) \leq (1-g_i),& i=1,...,n\\[5pt]
\label{Eq_ch_dars_dars_linear_c8}
&\text{ } 0 \leq y_i \leq 1,& i = 1,...,n\\[5pt]
\label{Eq_ch_dars_dars_linear_c9}
&\text{ } 0 \leq \theta_i \leq 1,& i = 1,...,n \\[5pt]
\label{Eq_ch_dars_dars_linear_c10}
& \text{ }x_i,g_i \in \{0,1\},& i=1,...,n
\end{align}

Convex optimization problems are solvable~\citep{boyd2004convex}. However, for problems of moderate to large sizes, integer linear programming (ILP) models are preferred~\citep{luenberger2015linear,mougouei2017integer} as they can be efficiently solved, despite the inherent complexity of NP-hard problems, due to the advances in solving ILP models and availability of efficient tools such as ILOG CPLEX for that purpose. This motivates us to consider developing an ILP version of the model as given by~(\ref{Eq_ch_dars_dars_linear}). 

In doing so, non-linear expression $x_i\theta_i$ is substituted by linear expression $y_i$ ($y_i=x_i\theta_i$). As such, either $a:(x_i=0,y_i=0)$, or $b:(x_i=1,y_i=\theta_i)$ occur. To capture the relation between $\theta_i$ and $y_i$ in a linear form, we have made use of an auxiliary variable $g_i=\{0,1\}$ and (\ref{Eq_ch_dars_dars_linear_c4})-(\ref{Eq_ch_dars_dars_linear_c8}) are added to the original model. As such, we have either $(g_i=0) \rightarrow a$, or $(g_i=1) \rightarrow b$. Therefore,~(\ref{Eq_ch_dars_dars_linear})-(\ref{Eq_ch_dars_dars_linear_c10}) is linear and can be efficiently solved~\citep{boyd2004convex}, even for large scale requirement sets, by existing commercial solvers such as \textit{IBM CPLEX}. 

%% file: case.tex
\section{Case Study}
\label{ch_dars_validation_case}

This section discusses the practicality and validity of DARS by studying a real-world software project. We demonstrate why software vendors should take care with value dependencies among requirements, and how to employ DARS to assist decision makers to comprehend the results, thus raising the following research questions.

\input{rq/rq_dars_ilp_case}

%% file: rq/rq_dars_ilp_case.tex
\begin{itemize}[leftmargin=1.5cm]
	\itemsep0em 
	\hypertarget{RQ1}{ }
	\item[(\textbf{RQ1})] How effective is DARS in considering value dependencies? 
	\hypertarget{RQ1.1}{ } 
	\item[(\textbf{RQ1.1})] How similar are the solutions found by DARS to those found by other selection methods?
	\hypertarget{RQ1.2}{ } 
	\item[(\textbf{RQ1.2})] What is the impact of using DARS on the overall value of software?  
	\hypertarget{RQ1.3}{ }
	\item[(\textbf{RQ1.3})] What is the relationship between maximizing the accumulated value, expected value, and overall value of a software product?
	\hypertarget{RQ1.4}{ }
	\item[(\textbf{RQ1.4})] How effective is DARS in mitigating value loss?
\end{itemize}

%% file: case_description.tex
\subsection{Description of Study}
\label{ch_dars_validation_case_description}

To demonstrate the practicality of DARS, we studied a real-world software project. Table~\ref{table_ch_dars_cost_value} lists the requirements of the project and their estimated and expected values in $[1,20]$. The expected value of each requirement $r_i$, denoted by $E(v_i)$ was computed by multiplying the frequency of the presence of $r_i$ in the configurations of the project sold in the earlier versions of software ($p(r_i)$) by its estimated value $v_i$. 

\begin{table*}[!htbp]
	\caption{The estimated and expected values of the requirements.}
	\label{table_ch_dars_cost_value}
	\centering
	\input{table_value}
\end{table*} 

\begin{figure*}[!htbp]
	\centering\includegraphics[scale=0.235]{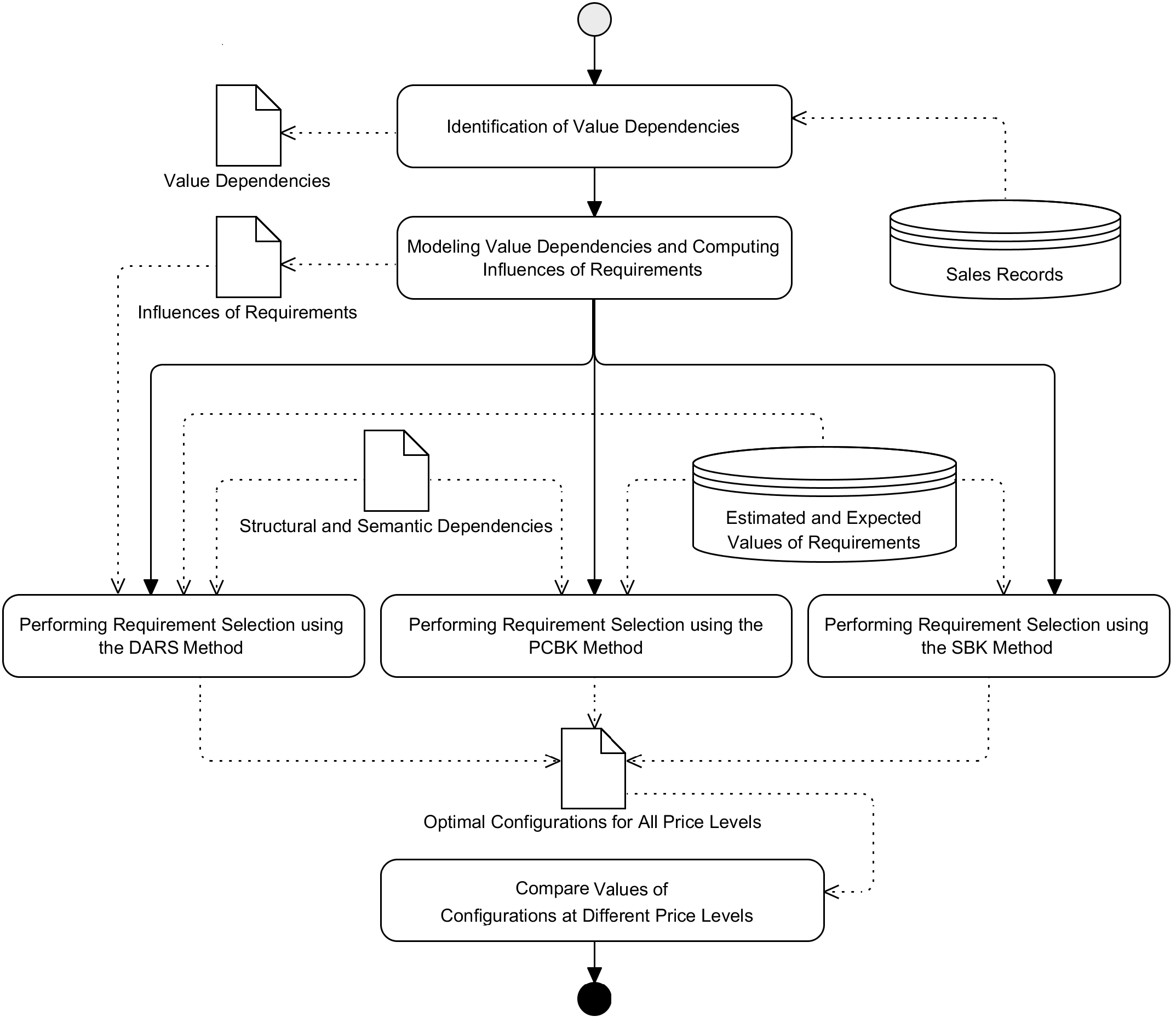}
	\captionsetup{margin=2ex}
	\caption{%
		The case study design.
	}%
	\label{fig_ch_dars_cs_design}
\end{figure*}

Our study began with the identification of value dependencies and modeling those dependencies as depicted in Figure~\ref{fig_ch_dars_cs_design}. Then we performed requirements selection using PCBK, SBK, and DARS based on the sales records of the previously released configurations of software. For the configurations found by PCBK, SBK, and DARS (Figure~\ref{fig_ch_dars_selectionMap_pcbk_dars} and Figure~\ref{fig_ch_dars_cs_selectionMap_sbk_dars}), the accumulated value (AV), expected value (EV), and overall value (OV) were computed to compare the performance of those methods for different price levels. This helped stakeholders to find, for different price levels, configurations with lower risk of value loss.  

%% file: table_value.tex
\scriptsize\resizebox {0.78\textwidth }{!}{
\begin{tabular}{llllllll}
	\toprule[1.5pt]
	\cellcolor[HTML]{FFFFFF}{\color[HTML]{000000} \textbf{$r_i$}} & \cellcolor[HTML]{FFFFFF}{\color[HTML]{000000} \textbf{$p(r_i)$}} & \cellcolor[HTML]{FFFFFF}{\color[HTML]{000000} \textbf{$v_i$}} & \cellcolor[HTML]{FFFFFF}{\color[HTML]{000000} \textbf{$E(v_i)$}} & \cellcolor[HTML]{FFFFFF}{\color[HTML]{000000} \textbf{$r_i$}} & \cellcolor[HTML]{FFFFFF}{\color[HTML]{000000} \textbf{$p(r_i)$}} & \cellcolor[HTML]{FFFFFF}{\color[HTML]{000000} \textbf{$v_i$}} & \cellcolor[HTML]{FFFFFF}{\color[HTML]{000000} \textbf{$E(v_i)$}} \\ \midrule

	$r_1$ & $00.94$ & $10.00$ & $09.43$ & $r_{15}$ & $00.58$ & 08.00 & 04.64 \\
	$r_2$ & $01.00$ & $20.00$ & $20.00$ & $r_{16}$ & $00.82$ & 10.00 & 08.24 \\
	$r_3$ & $00.37$ & $05.00$ & $01.85$ & $r_{17}$ & $00.12$ & 10.00 & 01.19 \\
	$r_4$ & $00.98$ & $17.00$ & $16.61$ & $r_{18}$ & $00.51$ & 15.00 & 07.59 \\
	$r_5$ & $00.88$ & 06.00 & 05.28 & $r_{19}$ & $00.67$ & 20.00 & 13.41 \\
	$r_6$ & $00.91$ & 20.00 & 18.30 & $r_{20}$ & $00.20$ & 20.00 & 04.09 \\
	$r_7$ & $00.82$ & 15.00 & 12.36 & $r_{21}$ & $00.14$ & 15.00 & 02.05 \\
	$r_8$ & $01.00$ & 09.00 & 09.00 & $r_{22}$ & $00.33$ & 20.00 & 06.59 \\
	$r_9$ & $00.97$ & 20.00 & 19.43 & $r_{23}$ & $00.88$ & 20.00 & 17.61 \\
	$r_{10}$ & $00.76$ & 16.00 & 12.18 & $r_{24}$ & $01.00$ & 01.00 & 01.00 \\
	$r_{11}$ & $00.57$ & 20.00 & 11.36 & $r_{25}$ & $00.24$ & 05.00 & 01.19 \\
	$r_{12}$ & $01.00$ & 12.00 & 12.00 & $r_{26}$ & $00.36$ & 01.00 & 00.36 \\
	$r_{13}$ & $00.76$ & 08.00 & 06.09 & $r_{27}$ & $00.97$ & 05.00 & 04.86 \\
	$r_{14}$ & $00.45$ & 14.00 & 06.28 &  &  &  &  \\[2pt] \midrule
	{\color[HTML]{000000} \textbf{Sum}} & {\color[HTML]{000000} -} & {\color[HTML]{000000} 192.00} & {\color[HTML]{000000} 160.17} & {\color[HTML]{000000} -} & {\color[HTML]{000000} -} & {\color[HTML]{000000} 150.00} & \cellcolor[HTML]{FFFFFF}{\color[HTML]{000000} 72.82} \\ \bottomrule[1.5pt]
\end{tabular}

}

%% file: case_identification.tex
\subsection{Identifying and Modeling Value Dependencies}
\label{ch_dars_case_identification}

\begin{figure*}[!htbp]
	\centering\includegraphics[scale=0.8]{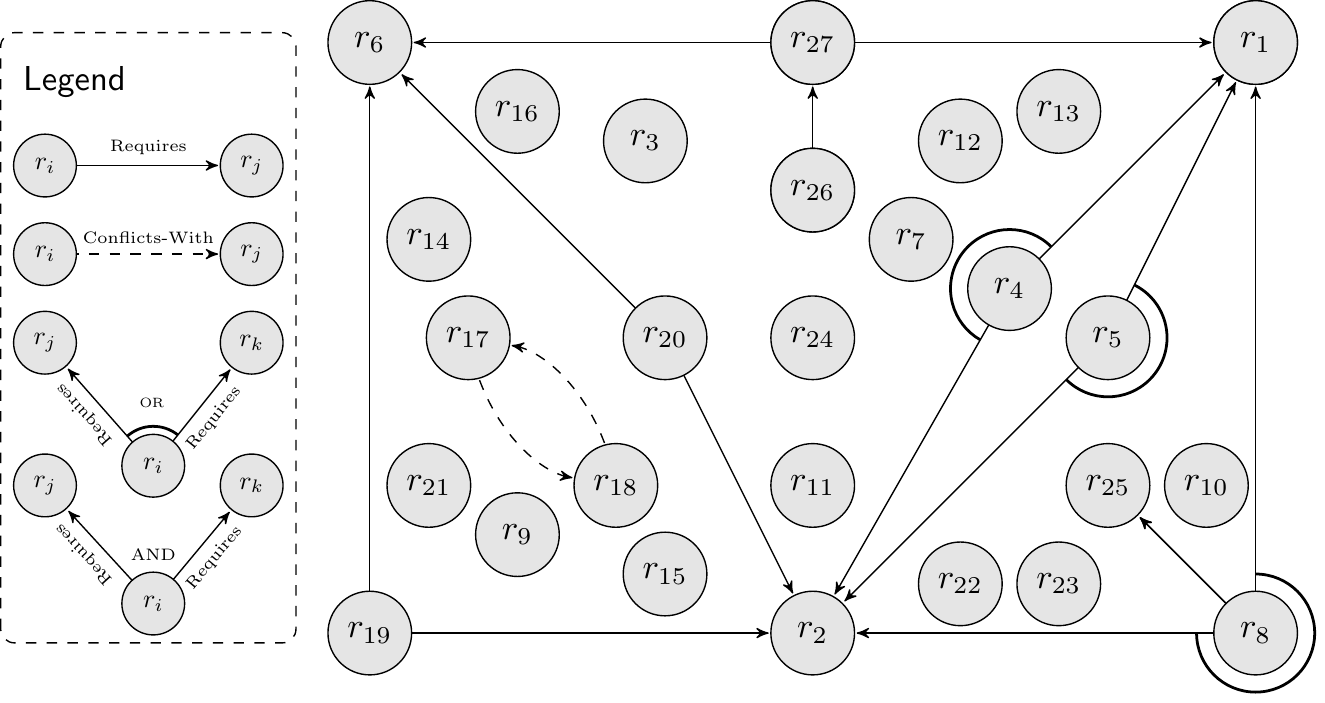}
	\vspace{-0.25cm}
	\caption{%
		The precedence dependency graph of the requirements.
	}%
	\label{fig_ch_dars_cs_precede}
\end{figure*}

To account for the precedence dependencies among the requirements of the project, dependencies of type \textit{Requires} and \textit{Conflicts-With} were extracted (Figure~\ref{fig_ch_dars_cs_precede}) from the development artifacts of the project and formulated as (\ref{eq_ch_dars_pcbk_cs_precedence_2})-(\ref{eq_ch_dars_pcbk_cs_precedence_14}) in the optimization models of PCBK, SBK, and DARS. Moreover, (\ref{eq_ch_dars_pcbk_cs_precedence_1}) was added to account fo the constraint that the presence of either $r_2$ or $r_6$ is always essential. $x_i$ denotes whether $r_i$ is selected ($x_i=1$) or not ($x_i=0$). 

\begin{align}
\label{eq_ch_dars_pcbk_cs_precedence_1}
&x_2 + x_6 =1\\
\label{eq_ch_dars_pcbk_cs_precedence_2}
&x_4\leq x_1 + x_2\\
\label{eq_ch_dars_pcbk_cs_precedence_3}
&x_5\leq x_1 + x_2\\
\label{eq_ch_dars_pcbk_cs_precedence_4}
&x_8\leq x_1 + x_2\\
\label{eq_ch_dars_pcbk_cs_precedence_5}
&x_8\leq x_{25}\\
\label{eq_ch_dars_pcbk_cs_precedence_6}
&x_{17}\leq (1-x_{18})\\
\label{eq_ch_dars_pcbk_cs_precedence_7}
&x_{18}\leq (1-x_{17})\\
\label{eq_ch_dars_pcbk_cs_precedence_8}
&x_{19}\leq x_2\\
\label{eq_ch_dars_pcbk_cs_precedence_9}
&x_{19}\leq x_6\\
\label{eq_ch_dars_pcbk_cs_precedence_10}
&x_{20}\leq x_2\\
\label{eq_ch_dars_pcbk_cs_precedence_11}
&x_{20}\leq x_6\\
\label{eq_ch_dars_pcbk_cs_precedence_12}
&x_{26}\leq x_{27}\\
\label{eq_ch_dars_pcbk_cs_precedence_13}
&x_{27}\leq x_{1}\\
\label{eq_ch_dars_pcbk_cs_precedence_14}
&x_{27}\leq x_{6}
\end{align}

To find value dependencies among the requirements of the project, we first collected sales records of different configurations of the project as explained earlier. Then the Eells measure of causal strength was computed for all pairs of requirements using Algorithm~\ref{ch_dars_alg_identification} to identify the strengths and qualities of causal relations among the requirements as explained in Section~\ref{ch_dars_identification_relations}. The significances of the identified relations were subsequently tested using the Odds Ratio at confidence level $95\%$ as explained in Section~\ref{ch_dars_identification_testing}. The strengths and qualities of explicit value dependencies were finally computed using the significant causal relations found and the fuzzy membership function of Figure~\ref{fig_ch_dars_membership_1} as given by~(\ref{Eq_ch_dars_strengthMeasure})-(\ref{Eq_ch_dars_qualityMeasure}). Algorithm~\ref{alg_ch_dars_strength} was used to infer implicit value dependencies and compute the overall strengths of positive and negative value dependencies in the value dependency graph (VDG) of the requirements. The influences of the requirements on the values of each other were then computed by (\ref{Eq_ch_dars_influence}). 

\begin{figure*}[!htbp]
	\centering
	\hspace{0.0cm}\includegraphics[scale=0.33]{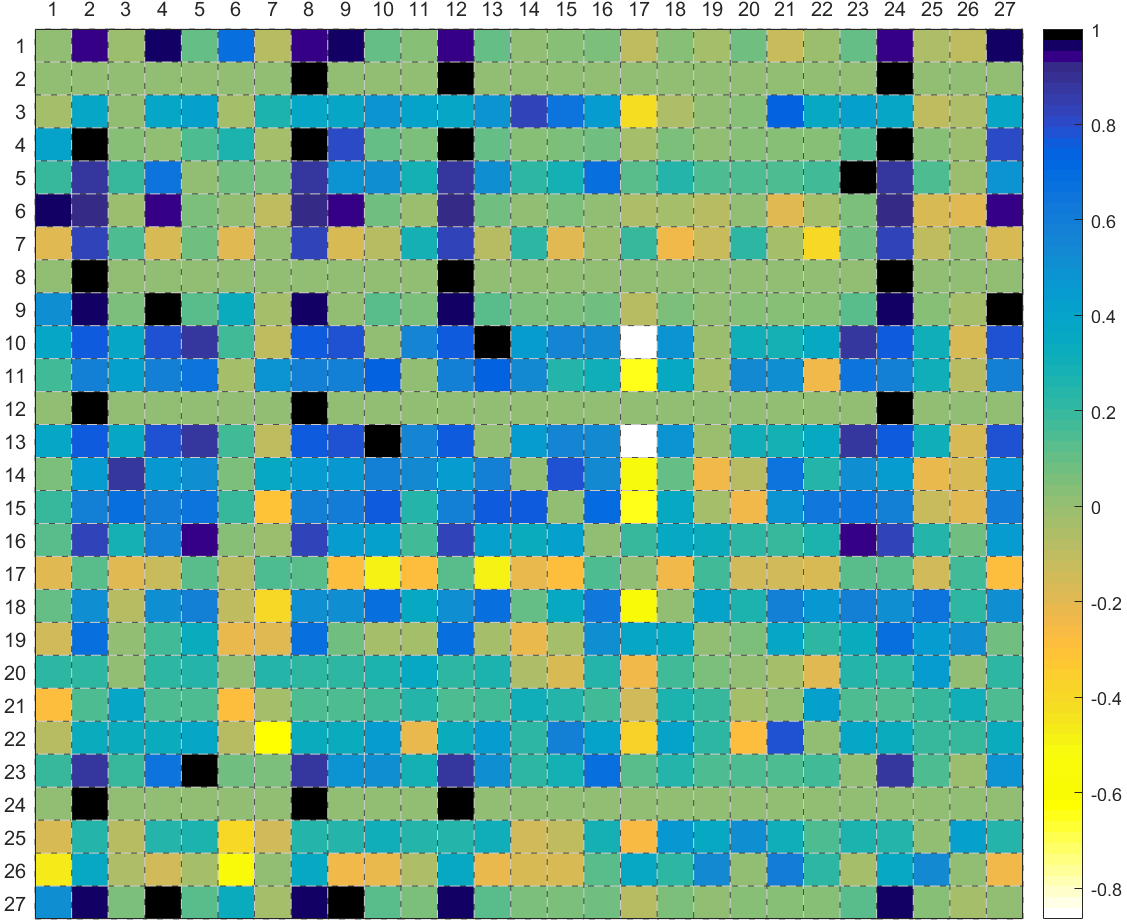}
	\captionsetup{}
	\caption{%
		Explicit value dependencies among the requirements. Row $i$ and column $j$ denotes quality and strength of a value dependency from requirement $r_i$ to $r_j$.
	}%
	\label{fig_ch_dars_dependencies}
\end{figure*}

\begin{figure*}[!htbp]
	\centering
	\hspace{0.0cm}\includegraphics[scale=0.33]{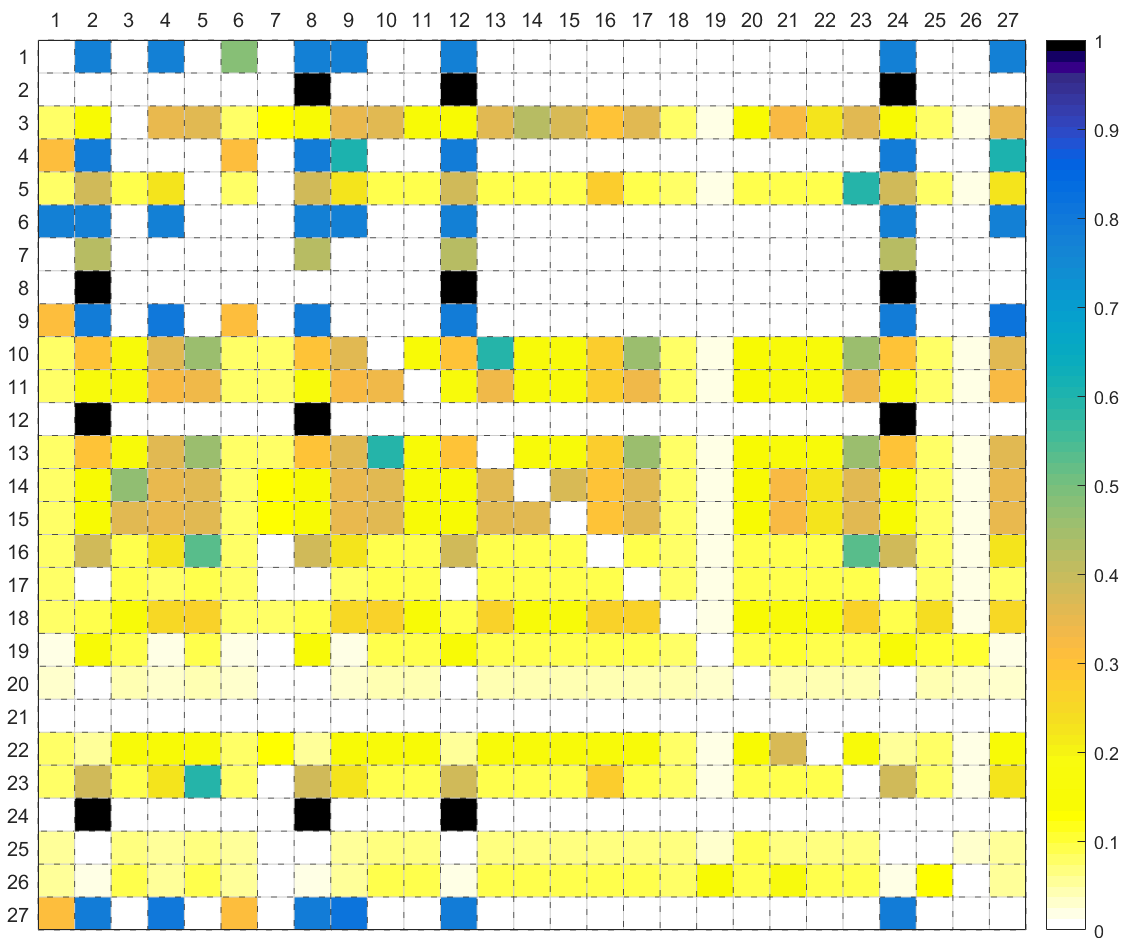}
	\captionsetup{}
	\caption{%
		Influences among the requirements: cell $i,j$ gives the influence of $r_j$ on $r_i$.
	}%
	\label{fig_ch_dars_influences}
\end{figure*}

Figure~\ref{fig_ch_dars_dependencies} shows the qualities and strengths of explicit value dependencies. The color of a cell at row $i$ and column $j$ specifies the quality and the strength of a value dependency from $r_i$ to $r_j$. Colors associated with positive (negative) numbers denote positive (negative) dependencies. Also, zero denotes the absence of any value dependency. Similarly, the positive or negative influences of the requirements on the values of each other are depicted in Figure~\ref{fig_ch_dars_influences}.


%% file: case_selection.tex
\subsection{Performing Requirements Selection}
\label{ch_dars_validation_case_selection}

This section demonstrates the effectiveness of DARS in considering value dependencies compared to BK, PCBK and SBK methods. As discussed in Section~\ref{sec_related_pcbk}, PCBK considers the estimated values of requirements while SBK, as in Section~\ref{sec_related_sbk}, accounts for user preferences for requirements by considering the expected values of requirements rather than their estimated values. Finally, DARS method factors in both user preferences and value dependencies among requirements as explained in Section~\ref{ch_dars_selection_ilp}. The expected values of the requirements and value dependencies among them are computed based on the user preferences achieved from the sales records of the project.  

\begin{align}
\label{eq_ch_dars_pcbk_cs_c1}
&\sum_{i=1}^{27} v_i  x_i \leq \gamma 
\end{align}

Price was determined as a major constraint for requirements selection as different configurations of the project had been released at different price levels earlier to cope with the needs of different users~\citep{karpoff1987relation}. Constraints (\ref{eq_ch_dars_pcbk_cs_c1}) hence was added to the optimization models of PCBK, SBK, and DARS respectively to contain the price of different configurations of software within their corresponding price limits. This converted the problem to a variation of Bounded Knapsack Problem. $\gamma$ $\in {\rm I\!R^+}$ denotes the price limit and $v_i$ specifies the estimated value of a requirement $r_i$. Also $x_i$ specifies whether a requirement $r_i$ is selected ($x_i=1$) or not ($x_i=0$). We omitted (\ref{eq_sbk_c1}) from the optimization model of SBK as considering the sales diversification is beyond the scope of this paper. The concept of diversification was explained in detail in Section~\ref{sec_related}. 

\begin{figure*}[!htbp]
    \centering
    \hspace{-1.1cm}\includegraphics[scale=0.52,angle=0]{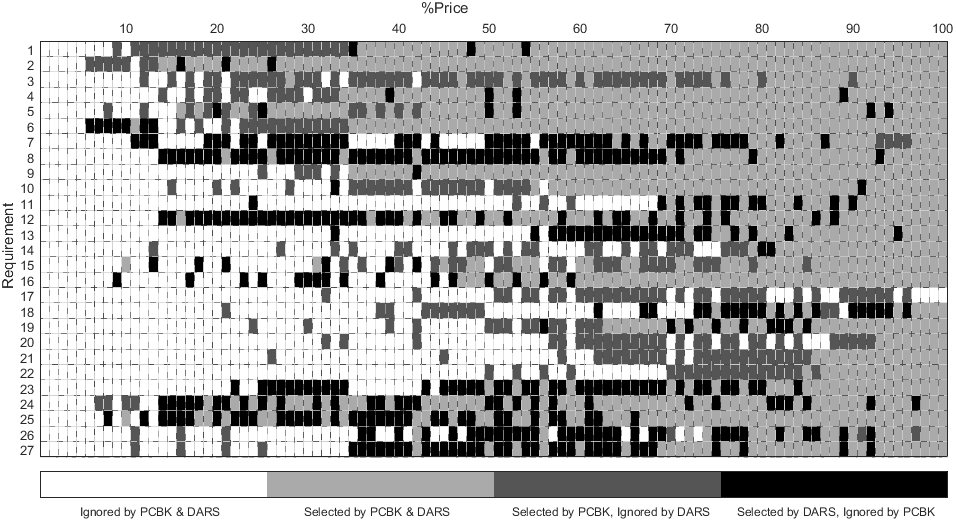}
    \caption{%
         Requirement subsets found by DARS and PCBK for different price levels. 
    }%
    \label{fig_ch_dars_selectionMap_pcbk_dars}
\end{figure*}

\begin{figure*}[!htbp]
    \centering
    \hspace{-1.1cm}\includegraphics[scale=0.52,angle=0]{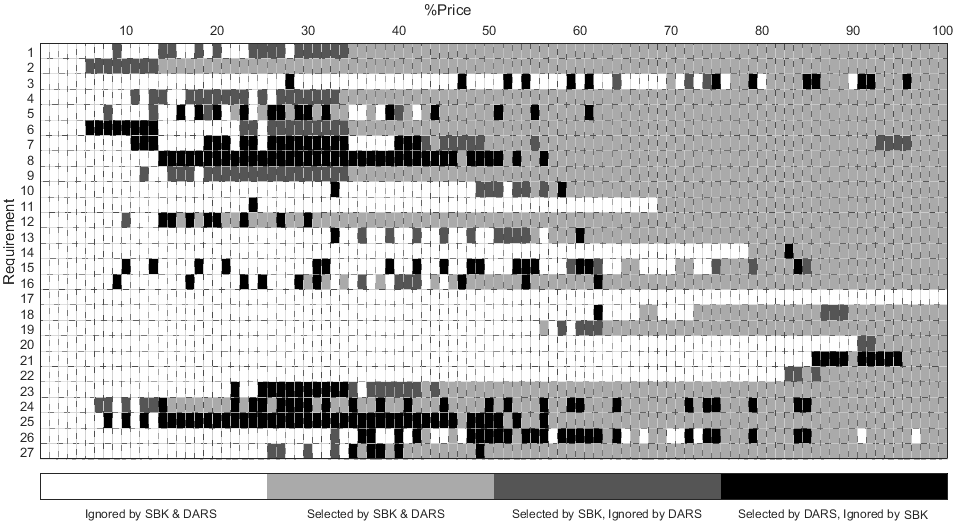}
    \caption{%
         Requirement subsets found by DARS and SBK for different price levels. 
    }%
    \label{fig_ch_dars_cs_selectionMap_sbk_dars}
\end{figure*}

Selection tasks were performed for different price levels ($\%\text{Price}=\{1,...,100\}$, $\text{Price} = \frac{\%\text{Price} }{100}\times 342$) using the optimization models of PCBK, SBK, and DARS with (\ref{eq_ch_dars_pcbk_cs_precedence_1})-(\ref{eq_ch_dars_pcbk_cs_precedence_14}) to find optimal subsets of the requirements (optimal configurations). Optimal configurations found by PCBK, SBK, and DARS were compared based on their similarities, accumulated values, expected values, and overall values to answer \protect\hyperlink{RQ1}{(\textbf{RQ1})} and its subquestions. Binary knapsack (BK) method (Section~\ref{sec_related_bk}) and the Increase-Decrease method (Section~\ref{sec_related_pcbk_CS}) were not used in the requirements selection tasks as the former ignores precedence dependencies resulting in violation of the precedence constraints while the latter does not provide any formal way to specify the amounts of the increased or decreased values of the requirement subsets as detailed in Section~\ref{sec_related_pcbk_CS}. Selections were performed using the callable library ILOG CPLEX 12.6.2 on a windows machine with a Core i7-2600 3.4 GHz processor and 16 GB of RAM.


\subsubsection{Similarities of the Solutions} 
\label{ch_dars_validation_case_selection_comparing}

\begin{figure*}[htbp]
    \begin{center}
        \subfigure[DARS vs. PCBK]{%
            \label{fig_distance_minkowski}
            \includegraphics[scale=0.46]{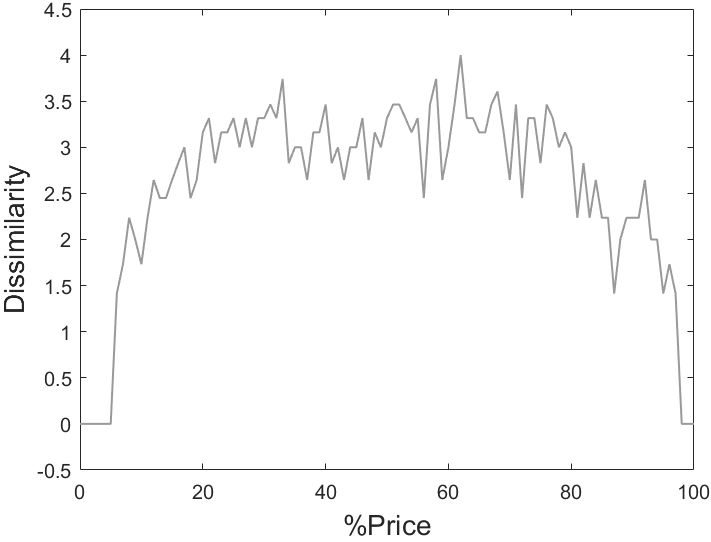}
        }
        \subfigure[DARS vs. SBK]{%
            \label{fig_distance_euclidean}
            \includegraphics[scale=0.46]{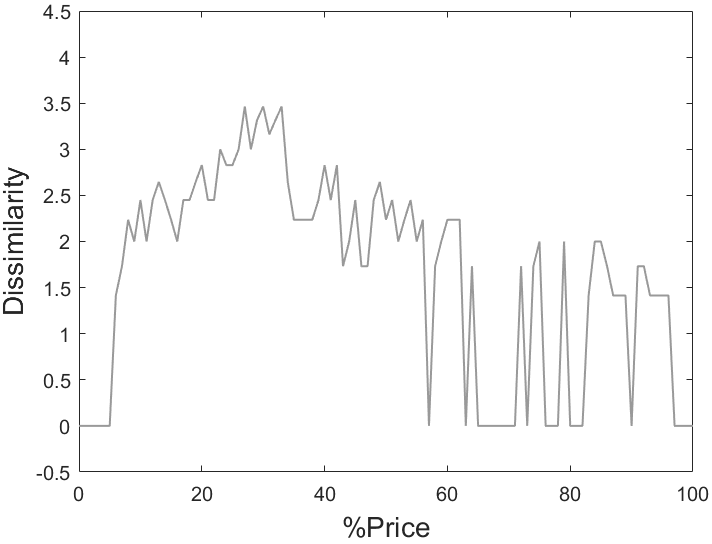}
        }
    \end{center}
    \captionsetup{margin=0ex}
    \vspace{-0.3cm}
    \caption{%
        Dissimilarities of the solutions found by DARS and PCBK/SBK. 
    }%
    \label{fig_distance}
\end{figure*}

In this section, we compare PCBK, SBK, and DARS based on their selection patterns to answer \protect\hyperlink{RQ1.1}{(\textbf{RQ1.1})}. Figure~\ref{fig_distance} depicts dissimilarities between the requirement subsets found by the DARS and those found by PCBK/SBK based on \textit{Euclidean Distance}. While notable at all price levels, these dissimilarities decreased for highly expensive ($\%\text{Price} \rightarrow 100$) or very cheap ($\%\text{Price} \rightarrow 0$) configurations of the project. The reason is expensive configurations of software comprise most requirements thus reducing the chances that requirements with positive influences on the values of the selected requirements are ignored.

Moreover, there are no negative influences among requirements (Figure~\ref{fig_ch_dars_influences}). Hence, similarities between solutions found by the DARS method and those found by PCBK and SBK increase for expensive configurations of software. For cheaper configurations, price constraint limits the solution space for PCBK, SBK, and DARS especially preventing the DARS method from utilizing its advantage in considering value dependencies. This resulted in more similarities between the solutions found by the DARS method and those found by PCBK and SBK for very cheap configurations of software. 

\begin{figure*}[htbp]
    \begin{center}
        \subfigure[DARS vs. PCBK]{%
            \label{fig_ch_dars_selection_dars_pcbk}
            \hspace{-0.35cm}\includegraphics[scale=0.57]{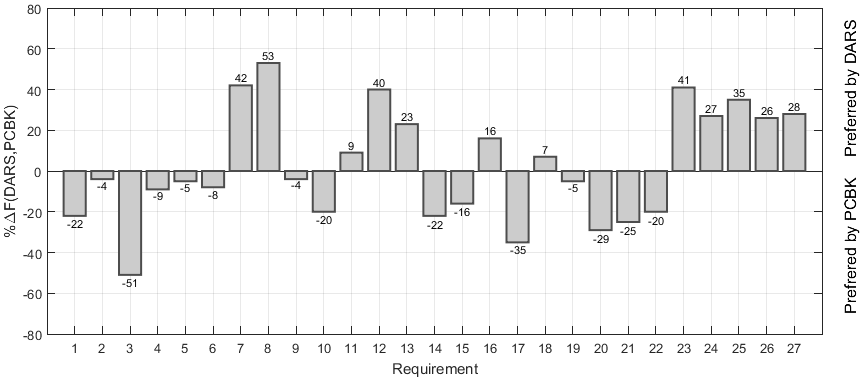}
        }
        \subfigure[DARS vs. SBK]{%
            \label{fig_ch_dars_selection_dars_sbk}
            \hspace{-0.35cm}\includegraphics[scale=0.57]{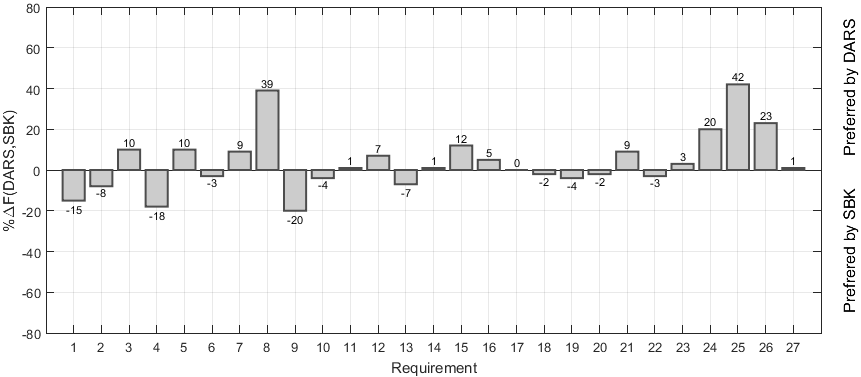}
        }
    \subfigure[SBK vs. PCBK]{%
        \label{fig_ch_dars_selection_sbk_pcbk}
        \hspace{-0.35cm}\includegraphics[scale=0.57]{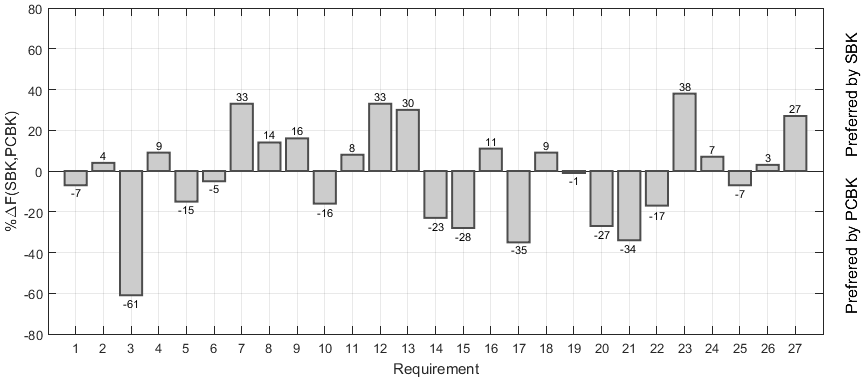}
    }
    \end{center}
    \captionsetup{margin=0ex}
    \caption{%
        Selection patterns of PCBK, SBK, and DARS at different price levels. For a requirement $r_i$, denoted by $i$ on the x-axis, and requirements selection methods $m_j$ and $m_k$, $\%\Delta \text{F}_{i}(m_j,m_k) = \%\text{F}_i (m_j)-\%\text{F}_i (m_k)$, where $\%\text{F}_i (m_j)$ and $\%\text{F}_i (m_k)$ give the percentage of the selection tasks in which $r_i$ is selected by $m_j$ and $m_k$ respectively. 
    }%
    \label{fig_ch_dars_selection}
\end{figure*}

Finally, we observed from Figure~\ref{fig_ch_dars_selection_dars_pcbk} and Figure~\ref{fig_ch_dars_selection_dars_sbk} that the requirement subsets (solutions) found by DARS were more similar to the solutions found by SBK than similar to the solutions found by PCBK. The reason is as explained before both DARS and SBK consider user preferences while PCBK ignores those preferences. 


Figure~\ref{fig_ch_dars_selection} provides more insights into \protect\hyperlink{RQ1.1}{(\textbf{RQ1.1})} by comparing the selection patterns of PCBK, SBK, and DARS in $100$ different selection tasks performed at different price levels ($\%\text{Price}=\{1,2,...,100\}$). For a given requirement $r_i$, $\%\text{F}_i (m_j)$ specifies the percentages of the selection tasks in which $r_i$ is selected by requirements selection method $m_j$. Hence, $\%\Delta \text{F}_{i}(m_j,m_k) = \%\text{F}_i (m_j)-\%\text{F}_i (m_k) >0 $ states that the percentages of the selection tasks where $r_i$ is selected by $m_j$ is higher than the percentages of the selection tasks where $r_i$ is selected by $m_k$. Similarly, $\%\Delta \text{F}_{i}(m_j,m_k) = \%\text{F}_i (m_j)-\%\text{F}_i (m_k) <0 $ states that $r_i$ is more frequently selected by $m_k$ compared to $m_j$. $m_j$ and $m_k$ can be any of the selection methods used in our selection tasks. 

We observed (Figure~\ref{fig_ch_dars_selection}) that requirements with significant influence on the values of pricey requirements were more frequently preferred by DARS compared to PCBK and SBK. This was more visible for requirements $r_8$, $r_{12}$, $r_{24}$, and $r_{27}$ when in Figure~\ref{fig_ch_dars_selection_dars_pcbk} and for requirements $r_8$, $r_{12}$, $r_{24}$ in Figure~\ref{fig_ch_dars_selection_dars_sbk}. 

Requirement $r_8$, for instance, was more frequently preferred by DARS compared to PCBK and SBK as the optimization model of DARS considers the fact that $r_8$ has a significant positive influence on the values of several valuable requirements including $r_2$, $r_4$, $r_6$, and $r_{12}$ (Figure~\ref{fig_ch_dars_influences}). Similarly, $r_{24}$ has a significant (positive) influence on the values of requirements $r_2$, $r_6$, $r_8$, and$r_{12}$. $r_{25}$ however, was more frequently selected by DARS as $r_8$ requires $r_{25}$ (Figure~\ref{fig_ch_dars_cs_precede}) and $r_8$ is frequently selected by DARS due to its significant impact on valuable requirements. As such, selecting $r_8$ requires the presence of $r_{25}$ in software. DARS and SBK however were frequently selected $r_{12}$ and $r_{27}$ as these two requirements are almost always preferred by users. This was not the case for PCBK as it ignores user preferences.   

Selection patterns in Figure~\ref{fig_ch_dars_selection} showed that when a decision was to be made regarding the presence or absence of a requirement $r_i$ in a configuration of software, PCBK only took into account the estimated value of $r_i$ ignoring user preferences. SBK on the other hand, considered user preferences for $r_i$ by evaluating the expected value of $r_i$ rather than merely its accumulated value. 

DARS, however, considered the expected value of $r_i$ and the impact of $r_i$ on the values of other requirements. More similarities were, therefore, observed among the configurations found by SBK and DARS as both methods took into account user preferences. On the contrary, dissimilarities were more visible when SBK and DARS/PCBK were compared as demonstrated in Figure~\ref{fig_ch_dars_selection_dars_pcbk} and Figure~\ref{fig_ch_dars_selection_sbk_pcbk}.   

\subsubsection{Impact of DARS on the Overall Value} 

\begin{figure*}[htbp]
    \begin{center}
        \subfigure[]{%
            \label{fig_ch_dars_ov_all}
            \includegraphics[scale=0.46]{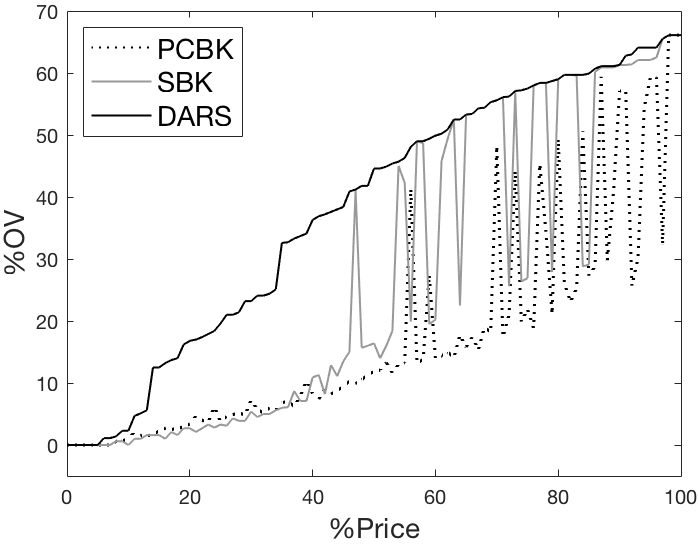}
        }
        \subfigure[]{%
            \label{fig_ch_dars_ov_delta_dars_pcbk}
            \includegraphics[scale=0.46]{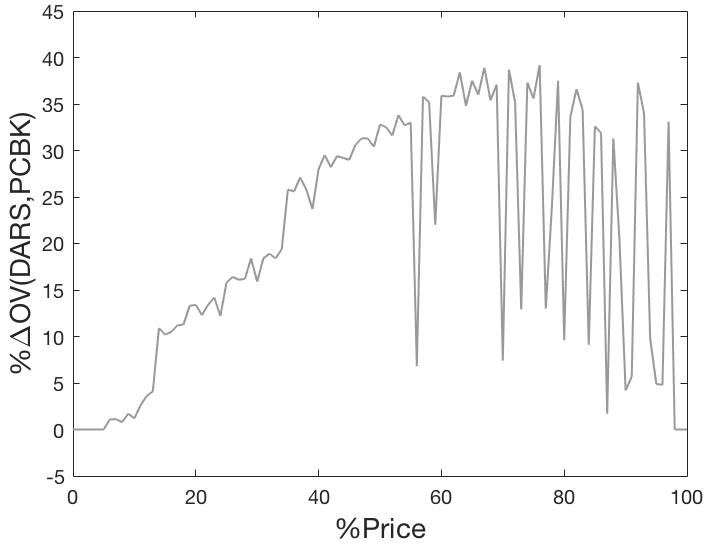}
        }
        \subfigure[]{%
            \label{fig_ch_dars_ov_delta_dars_sbk}
            \includegraphics[scale=0.46]{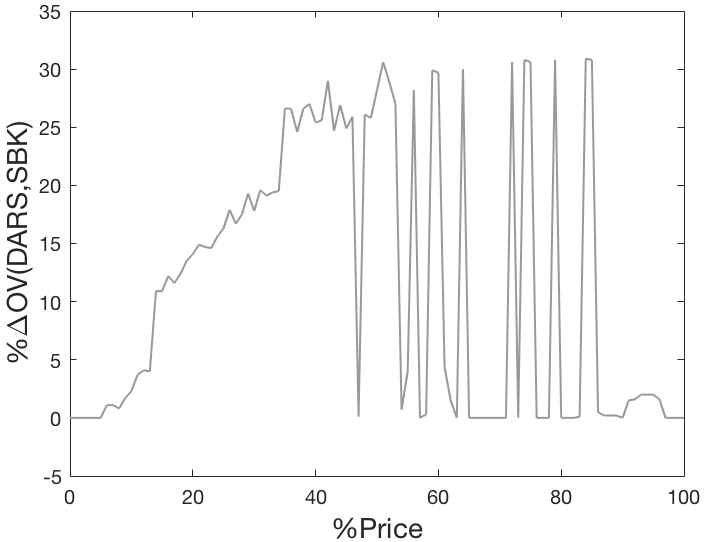}
        }
        \subfigure[]{%
            \label{fig_ch_dars_ov_delta_sbk_pcbk}
            \includegraphics[scale=0.46]{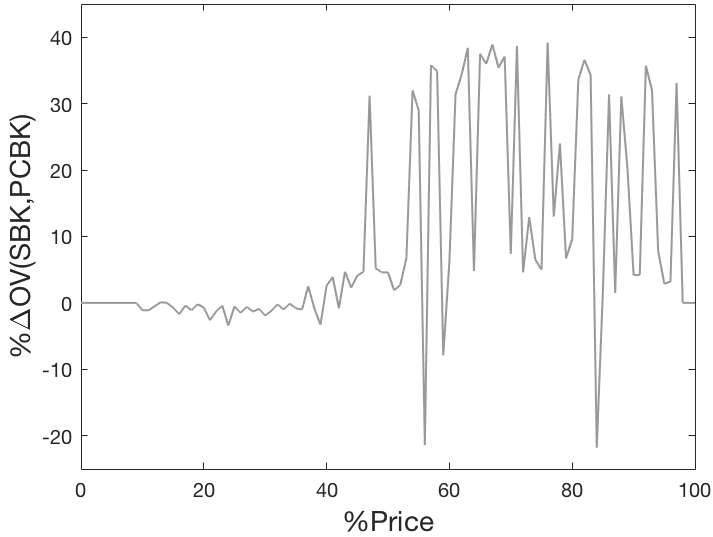}
        }
    \end{center}
    \captionsetup{margin=0ex}
    \vspace{-0.5cm}
    \caption{%
        Comparing the overall values at different price levels. $\%\Delta \text{OV}(m_j,m_k) = \%\text{OV}(m_j)-\%\text{OV}(m_k)$, where $m_j$ denotes a selection method (DARS, PCBK, or SBK).  
    }%
    \label{fig_ch_dars_ov}
\end{figure*}

\begin{figure*}[!htbp]
    \begin{center}
        \subfigure[]{%
            \label{fig_ch_dars_av_all}
            \includegraphics[scale=0.459]{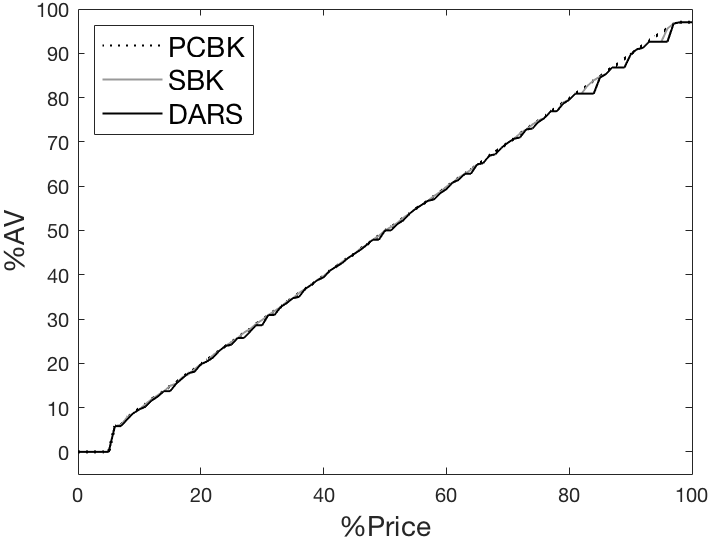}
        }
        \subfigure[]{%
            \label{fig_ch_dars_av_delta_dars_pcbk}
            \includegraphics[scale=0.459]{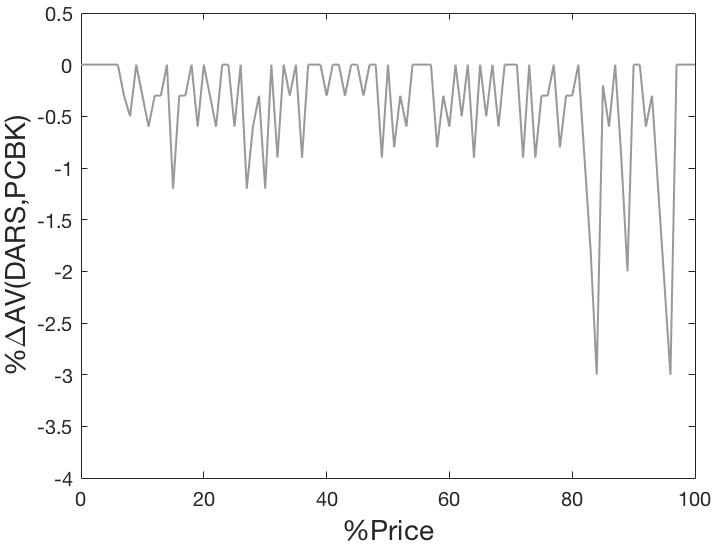}
        }
        \subfigure[]{%
            \label{fig_ch_dars_av_delta_dars_sbk}
            \includegraphics[scale=0.456]{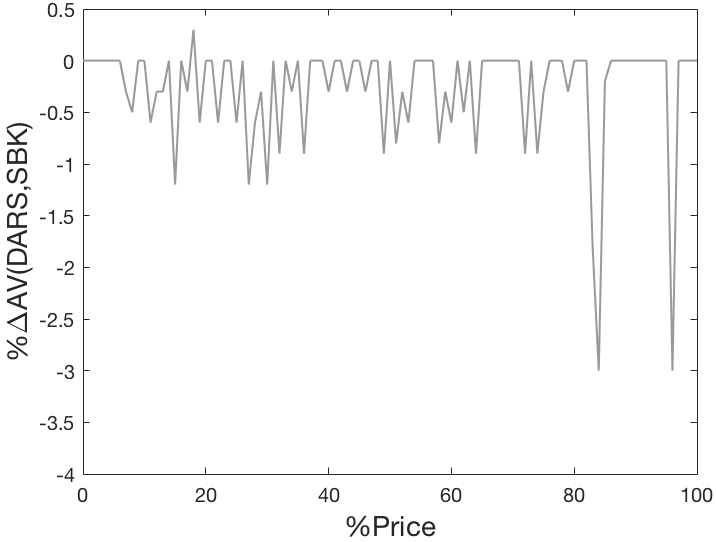}
        }
        \subfigure[]{%
            \label{fig_ch_dars_av_delta_sbk_pcbk}
            \includegraphics[scale=0.456]{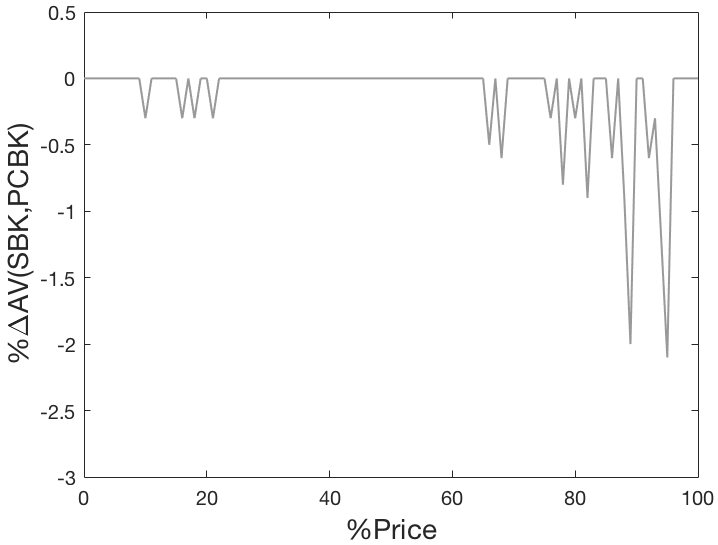}
        }
    \end{center}
    \captionsetup{margin=0ex}
    \vspace{-0.5cm}
    \caption{%
        Comparing the accumulated values at different price levels. $\%\Delta \text{AV}(m_j,m_k) = \%\text{AV}(m_j)-\%\text{AV}(m_k)$, where $m_j$ denotes a selection method (DARS, PCBK, or SBK).  
    }%
    \label{fig_ch_dars_av}
\end{figure*}

\begin{figure*}[htbp]
    \begin{center}
        \subfigure[]{%
            \label{fig_ch_dars_ev_all}
            \includegraphics[scale=0.46]{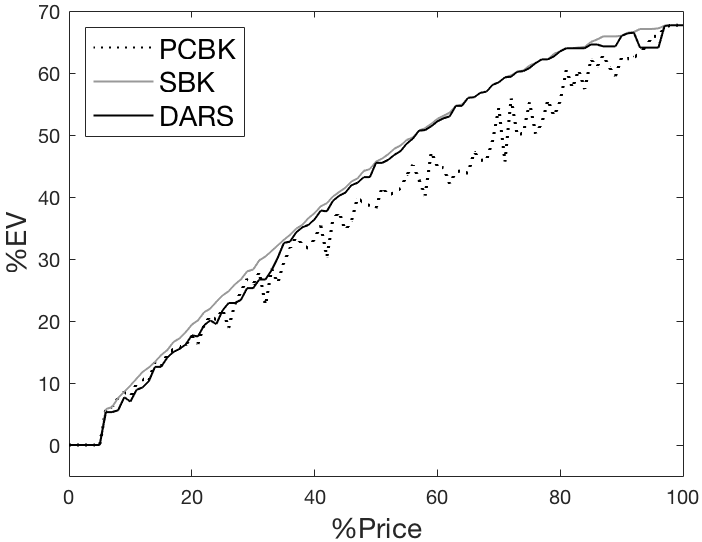}
        }
        \subfigure[]{%
            \label{fig_ch_dars_ev_delta_dars_pcbk}
            \includegraphics[scale=0.46]{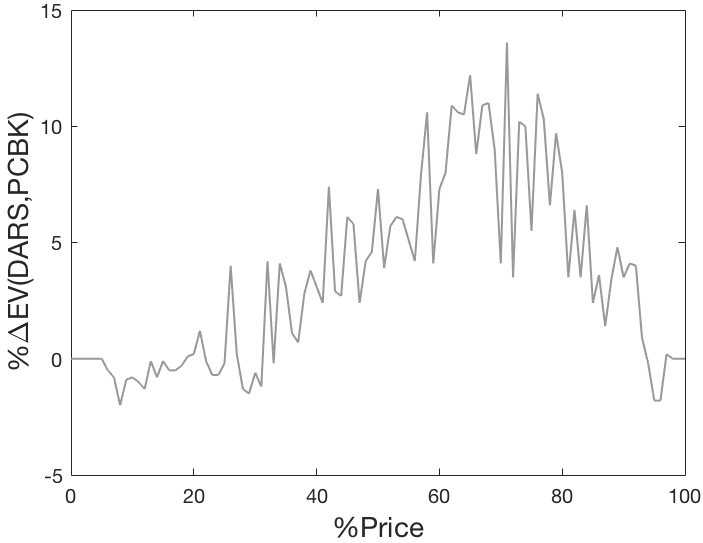}
        }
        \subfigure[]{%
            \label{fig_ch_dars_ev_delta_dars_sbk}
            \includegraphics[scale=0.46]{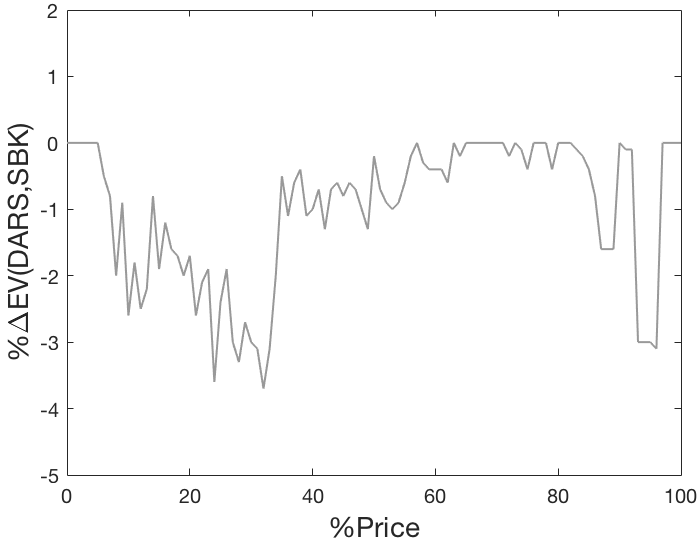}
        }
        \subfigure[]{%
            \label{fig_ch_dars_ev_delta_sbk_pcbk}
            \includegraphics[scale=0.46]{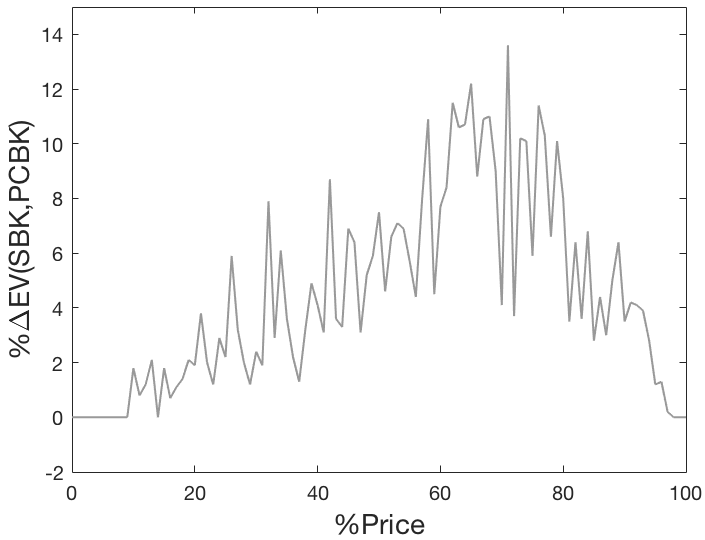}
        }
    \end{center}
    \captionsetup{margin=0ex}
    \caption{%
        Comparing the accumulated values at different price levels. $\%\Delta \text{EV}(m_j,m_k) = \%\text{EV}(m_j)-\%\text{EV}(m_k)$, where $m_j$ denotes a selection method (DARS, PCBK, or SBK).  
    }%
    \label{fig_ch_dars_ev}
\end{figure*}

\protect\hyperlink{RQ1.2}{(\textbf{RQ1.2})} is answered by comparing the percentages of overall values ($\%\text{OV}=(\text{OV}/342)\times 100$), accumulated values ($\%\text{AV}=(\text{AV}/342)\times 100$), and estimated values ($\%\text{EV}=(\text{EV}/342)\times 100$) provided by PCBK, SBK, and DARS for $100$ selection tasks, each performed at a specific price level ($\%\text{Price}=\{1,2,..,100\}$), as shown in Figures \ref{fig_ch_dars_ov}-\ref{fig_ch_dars_ev}. Our results show (Figure~\ref{fig_ch_dars_ov}) that requirement subsets found by DARS provided higher or equal $\%$OV in all selection tasks compared to the PCBK method. The reason is that the optimization model of PCBK ignores user preferences and value dependencies among the requirements.

For a given price, $\%OV$ of the requirement subset (solutions) found by the SBK method was, for most price levels, higher than $\%OV$ of the solution provided by the PCBK method but still less than or equal to the overall value of the solution found by DARS. The reason is that even though the SBK method does not consider value dependencies, it still accounts for user preferences, similar to DARS, by optimizing the expected values of selected requirements. This results in more similarities between the configurations found by the SBK method and those found by DARS as discussed in Section~\ref{ch_dars_validation_case_selection_comparing}. 

We observed in Figure~\ref{fig_ch_dars_ov_delta_dars_pcbk} that the gap between the $\%$OV achieved from DARS and the PCBK/SBK method was notable in almost all selection tasks performed at different price levels. But the gap reduced to almost negligible for highly expensive ($\%\text{Price} \rightarrow 100$) or very cheap ($\%\text{Price} \rightarrow 0$) configurations. The reason is, on one hand, there are no negative influences among the requirements (Figure~\ref{fig_ch_dars_influences}) and, on the other hand, expensive configurations of software comprise most requirements, which reduces the chances that requirements with positive influence are ignored by PCBK/SBK. 

That increases similarities between the expensive configurations found by PCBK/SBK and DARS as discussed in Section~\ref{ch_dars_validation_case_selection_comparing}. For cheaper configurations, the price-constraint limited the solution space in all the PCBK, SBK, and DARS and specially prevented DARS from utilizing its advantage in considering value dependencies. The price constraint further reduced the gap between $\%$AV provided by DARS and PCBK/SBK (Figures~\ref{fig_ch_dars_av}) in the selection tasks.

We further, observed insignificant differences amongst the accumulated values provided by the selection methods experimented in this study as shown in Figure~\ref{fig_ch_dars_av}. The reason is that the price constraint in the optimization models of the PCBK, SBK, and DARS contain the accumulated values of the solutions found by those models. The price constraint is needed to factor out the interplay between the price and sales as explained earlier. Moreover, the expected values of the requirement subsets found by the SBK method were higher than those found by DARS and PCBK in all selection tasks (for all price levels) as shown in Figure~\ref{fig_ch_dars_ev_delta_dars_sbk} and Figure~\ref{fig_ch_dars_ev_delta_sbk_pcbk} respectively. 

The expected values of requirement subsets found by DARS were higher than those of the requirement subsets found by the PCBK method in most selection tasks (Figure~\ref{fig_ch_dars_ev_delta_dars_pcbk}). In some of the selection tasks, however, the expected values of the requirement subsets found by the PCBK method were higher than those found by DARS even though the PCBK method does not account for user preferences. The reason is that DARS optimizes the overall value of a requirement subset (solution), which accounts for both user preferences and value dependencies. Hence in some cases DARS may find solutions with lower expected values as taking into account value dependencies may be in conflict with maximizing the expected values of a requirement subset.

\subsubsection{Understanding the Conflicting Objectives} 

To answer \protect\hyperlink{RQ1.3}{(\textbf{RQ1.3})}, we compared the overall values (Figure~\ref{fig_ch_dars_ov}), accumulated values (Figure~\ref{fig_ch_dars_av}), and expected values (Figure~\ref{fig_ch_dars_ev}) of the requirement subsets found by the PCBK, SBK, and DARS in different requirements selection tasks performed at different price levels. From Figure~\ref{fig_ch_dars_ov} and Figure~\ref{fig_ch_dars_av} it can be seen that maximizing the accumulated value (AV) of a selected subset of requirements conflicts with maximizing the overall value (OV) of that subset. This can be specially seen in Figure~\ref{fig_ch_dars_ov_delta_dars_pcbk} and Figure~\ref{fig_ch_dars_av_delta_dars_pcbk}, where in several selection tasks, choosing requirement subsets with higher $\%$AV by the PCBK method (Figure~\ref{fig_ch_dars_av}) reduced the overall value. 

Maximizing the expected value of a requirement subset also conflicts with optimizing its overall value as the former may result in ignoring requirements with lower expected values even if they have a significant influence on the values of other requirements. That will increase the penalty of ignoring requirements with positive influences on the values of selected requirements, as given by (\ref{eq_ch_dars_penalty}), resulting in lower overall value. This can be seen by comparing Figure~\ref{fig_ch_dars_ov_delta_dars_sbk} and Figure~\ref{fig_ch_dars_ev_delta_dars_sbk}.

\subsubsection{Mitigating the Value Loss} 

\begin{figure*}[!htbp]
    \centering
    \hspace{0.0cm}\includegraphics[scale=0.6]{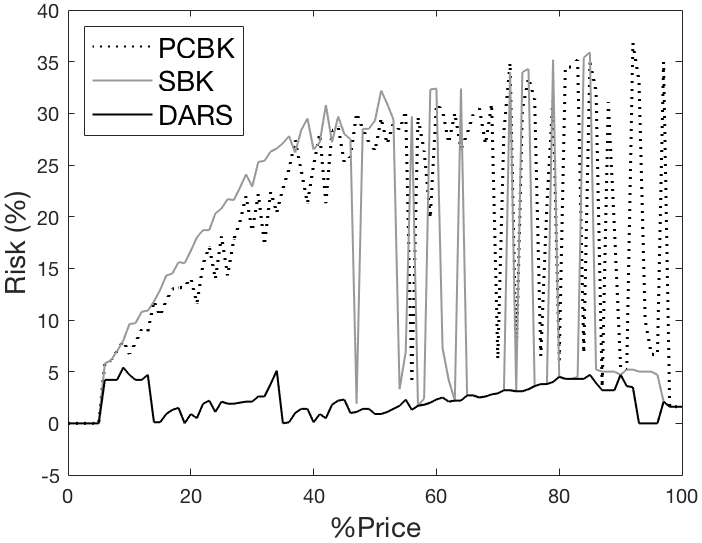}
    \caption{%
        Risk of value loss at different price levels. 
    }%
    \label{fig_ch_dars_risk}
\end{figure*}

Ignoring value dependencies can pose a risk to the economic worth of software configurations and eventually result in value loss as given by (\ref{eq_ch_dars_penalty}). This risk can be be measured by the gap between the expected value of software and its overall value, which accounts for value dependencies, as depicted in Figure~\ref{fig_ch_dars_risk}. As shown in this figure, for each selection task performed at a specific price level, the gap between the expected value and the overall value of software configuration found by DARS was notably smaller than the gaps between the $\%$EV and $\%$OV provided by the PCBK method. Hence, using DARS contributed to a smaller risk of value loss in different configurations of software. 

We observed (Figure~\ref{fig_ch_dars_risk}) that the risk of value loss for the configurations found by DARS was under $5\%$ while this fluctuated from almost negligible to around $37\%$ in PCBK and SBK. In most configurations found by PCBK and SBK, an inconsistent pattern of ``the higher the price the higher the risk of value loss" was observed suggesting a higher risk for expensive configurations of software. The risk of value loss for the configurations found by the SBK method, however, converged to those found by DARS for $\%\text{Price} \geq 88$ as both methods chose more similar configurations (\protect\hyperlink{RQ1.4}{(\textbf{RQ1.4})}).

%% file: scalability.tex
\section{Complexity and Scalability Analysis}
\label{ch_dars_scalability}
This section evaluates the scalability of DARS for identification and modeling value dependencies as well as considering those dependencies in software requirements selection. We specially generate random datasets with different numbers of requirements (up to $ 3000 $) to investigate the scalability of the ILP model of DARS for different scenarios in relation to value and precedence dependencies among requirements. Simulations thus were designed to answer the following questions. 

\input{rq/rq_dars_ilp_scalability}

\subsection{The Overhead of using DARS}

Our proposed DARS method relies on the identification and modeling of value dependencies -- that constitutes the main overhead of DARS. Identification of value dependencies from causal relations among user preferences is automated in DARS as explained in Section~\ref{ch_dars_identification}. The process, nevertheless, relies on computing the Eells measure~\citep{eells1991probabilistic} for pairs of the requirements. Algorithm~\ref{alg_ch_dars_strength} computes the Eells measure in $O(t \times n^2)$ for $n$ requirements and $t$ records of user preferences. Precedence dependencies among requirements (requires, conflicts-with, AND, OR) on the other hand, are identified as part of the requirement analysis and inferred from the structure and/or semantic of a software product using automated or semi-automated techniques~\citep{zhang2005feature,dahlstedt2005requirements}. This is an inevitable aspect of software requirement analysis and is not specific to DARS. Moreover, construction of a value dependency graph of requirements, inferring implicit value dependencies, and computing the influences of requirements using Algorithm~\ref{alg_ch_dars_strength} is of the computational complexity of $O(n^3)$ as discussed earlier in Section~\ref{ch_dars_modeling}. This concluded our answer to \protect\hyperlink{RQ2}{(\textbf{RQ2})}. 

\subsection{Scalability of the Optimization Model of DARS}

The optimization model of the DARS method as given by~(\ref{Eq_ch_dars_dars_linear})-(\ref{Eq_ch_dars_dars_linear_c8}) is scalable to datasets with a large number of requirements, different budget constraints, and various degrees of precedence/value dependencies. To demonstrate this, runtime simulations in Table~\ref{table_sim_scalability_design} were carried out. To simulate value dependencies for a desired VDL and NVDL, uniformly distributed random numbers in $[-1,1]$ were generated, where the sign and magnitude of each number specified the quality and the strength of its corresponding explicit value dependency. We used \textit{Precedence Dependency Level} (PDL) and \textit{Negative Precedence Dependency Level} (NPDL) as given by  (\ref{Eq_pdl}) and (\ref{Eq_npdl}) to specify the degree of precedence dependencies in a precedence graph $G$ with $n$ nodes (requirements). $k$ gives the total number of precedence dependencies while $j$ denotes the number of negative precedence dependencies in (\ref{Eq_pdl}) and (\ref{Eq_npdl}) respectively. 

\begin{table*}[!htbp]
	\caption{Runtime Simulations for the optimization model of DARS}
	\label{table_sim_scalability_design}
	\centering
	\input{table_sim_scalability_design}
\end{table*}

\begin{align}
\label{Eq_pdl}
&PDL(G)=\frac{k}{\Perm{n}{2}}=\frac{k}{n (n-1)} \\
\label{Eq_npdl}
&NPDL(G)=\frac{j}{k}
\end{align}

For a given PDL and NPDL, random numbers in $\{-1,0,1\}$ were generated where $1$ ($-1$) specified a positive (negative) precedence dependency and $0$ denoted the absence of any precedence dependency from a requirement $r_i$ to $r_j$. Simulations were carried out using the callable library ILOG CPLEX 12.6.2 on a windows machine with a Core i7-2600 3.4 GHz processor and 16 GB of RAM.

\begin{figure*}[htbp]
	\begin{center}
		\subfigure[Simulation 1]{%
			\label{fig_ch_dars_t_n_log}
			\includegraphics[scale=0.455]{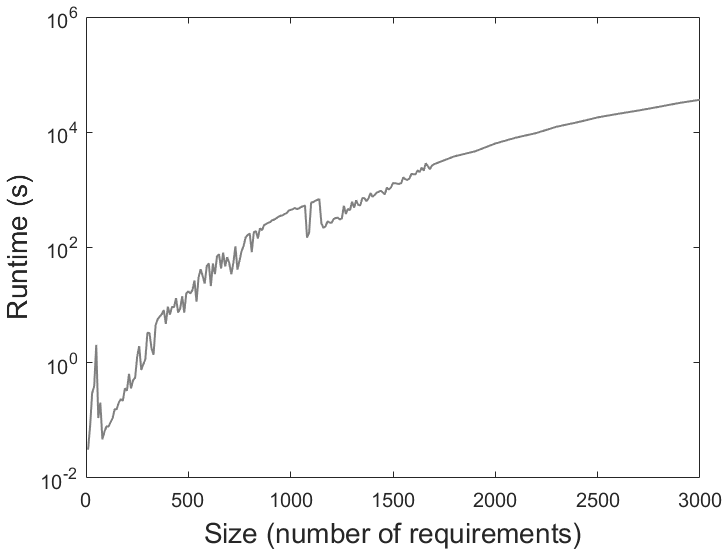}
		}
		\subfigure[Simulation 2]{%
			\label{fig_ch_dars_t_b}
			\includegraphics[scale=0.455]{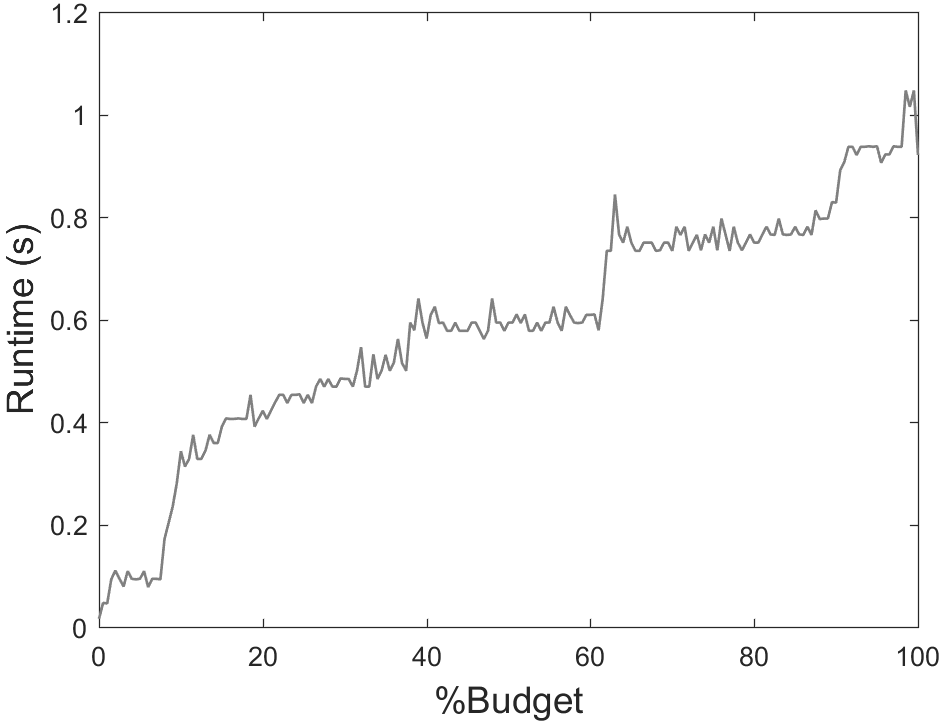}
		}
		\subfigure[Simulation 3]{%
			\label{fig_ch_dars_t_pdl}
			\includegraphics[scale=0.455]{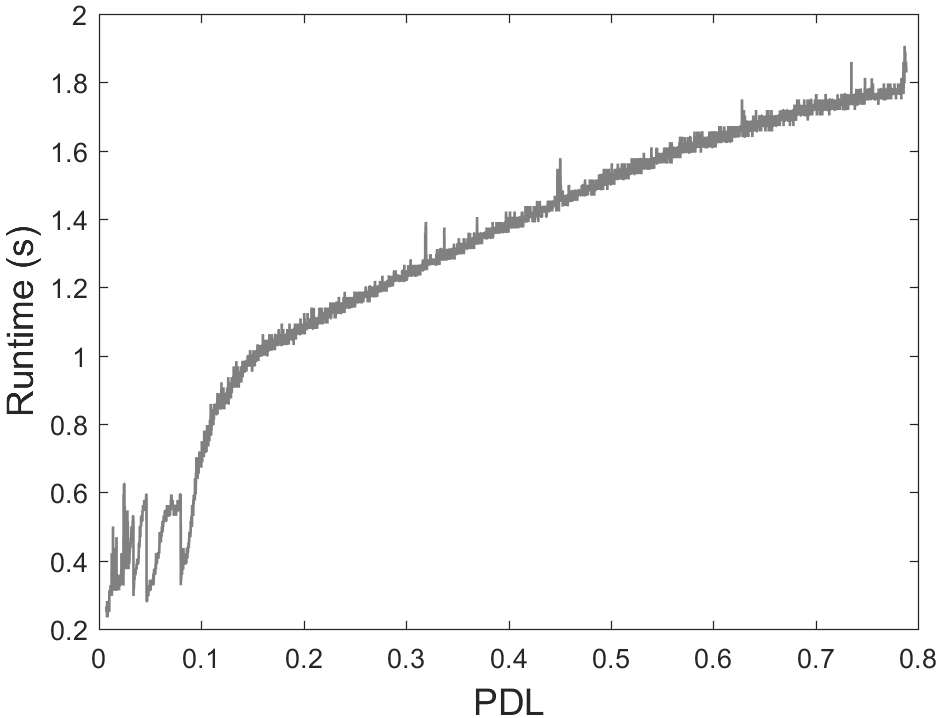}
		}
		\subfigure[Simulation 4]{%
			\includegraphics[scale=0.455]{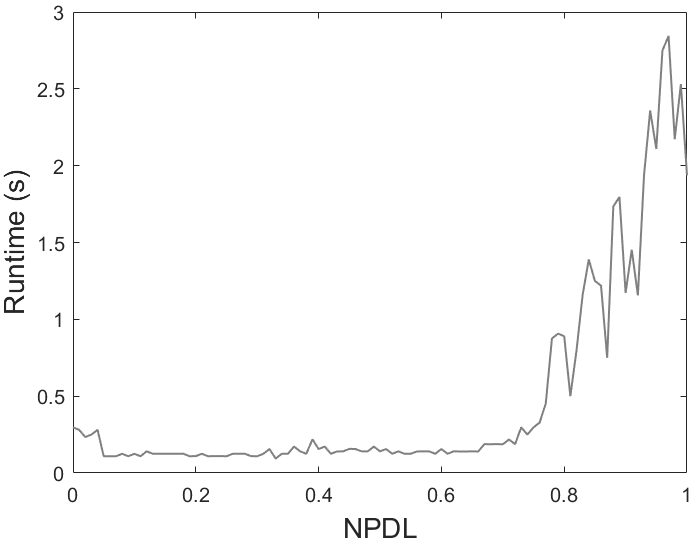}
			\label{fig_ch_dars_t_pdlnpdl}
		}
		\subfigure[Simulation 5]{%
			\includegraphics[scale=0.455]{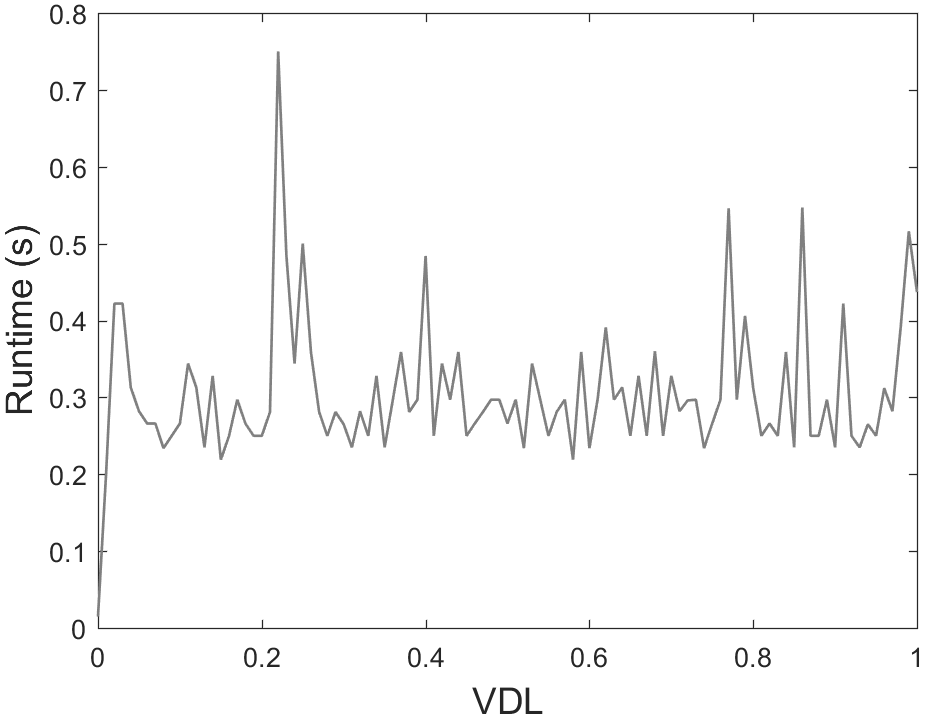}
			\label{fig_ch_dars_t_vdl}
		}
		\subfigure[Simulation 6]{%
			\includegraphics[scale=0.455]{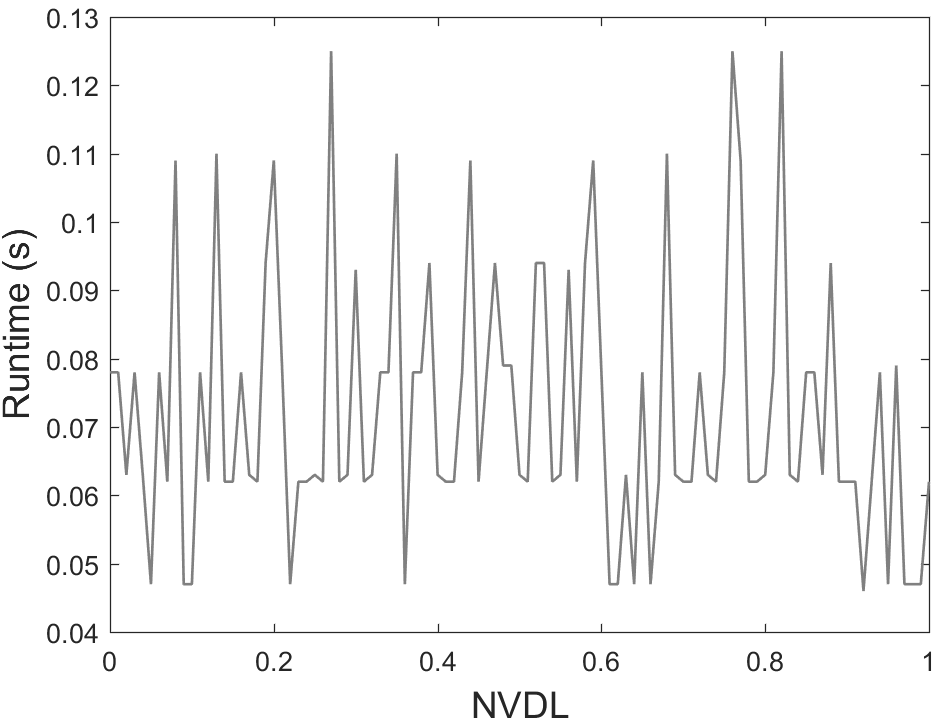}
			\label{fig_ch_dars_t_nvdl}
		}
	\end{center}
	\captionsetup{margin=5ex}
	\caption{%
		Runtime of DARS.
	}%
	\label{fig_ch_dars_ev}
\end{figure*}

\protect\hyperlink{RQ3.1}{(\textbf{RQ3.1})} is answered by runtime simulation 1, which evaluates the runtime of the optimization model of DARS for different numbers of requirements (Figure~\ref{fig_ch_dars_t_n_log}). We observed that increasing the number of requirements increased, as expected, the runtime of the optimization model of the DARS method. Nonetheless, for requirement sets with up to $750$ ($n\leq 750$) requirements, the model managed to find the optimal solution in less than a minute. For $750 < n\leq 2000$ the runtime was above one minute but did not exceed two hours. Finally, for $2000 < n \leq  3000$ it took hours before the selection was completed. On the other hand, our results for Simulation 2 demonstrated (Figure~\ref{fig_ch_dars_t_b}) that the runtime of the optimization model of the DARS method increased with a budget increase. The reason is with more budget, more requirements can be selected which results in a larger solution space.

As such, it may take longer for the optimization model of DARS to find the optimal subset. This answers \protect\hyperlink{RQ3.2}{(\textbf{RQ3.2})}. To answer \protect\hyperlink{RQ3.3}{(\textbf{RQ3.3})}, we simulated requirements selection for various precedence dependency levels (PDLs). Our results (Figure~\ref{fig_ch_dars_t_pdl}) demonstrated, that, in general, the runtime of the optimization model of DARS increased when PDL increased. The reason is increasing PDL limits the number of choices for the optimization model of the DARS method as the model needs to respect precedence dependencies; it takes longer for the selection task to complete. Increasing NPDL, on the other hand, had no significant impact on the runtime of the optimization model of DARS in most places. Nonetheless, for larger NPDLs ($NPDL \rightarrow 1$), runtime was increased. The reason is at such high NPDL, the optimization model of DARS cannot find a feasible solution with some values as each requirement conflicts with almost every other requirement. Hence, it takes longer for the optimization to complete and return the null set ($\%$OV=0) as the only feasible solution.    

Simulation 5 was carried out to answer \protect\hyperlink{RQ3.4}{(\textbf{RQ3.4})} by measuring the runtime of the selection models in the presence of various value dependency levels (VDLs). Our results demonstrate (Figure~\ref{fig_ch_dars_t_vdl}) that increasing (decreasing) VDL has an inconsistent impact of negligible magnitude on the runtime of the optimization model of the DARS method. In a similar way, our simulations for various negative value dependency levels (NVDLs) showed (Figure~\ref{fig_ch_dars_t_nvdl}) that the impact of increasing (decreasing) NVDL on the runtime of the optimization model of the DARS method was unpredictable. 

%% file: rq/rq_dars_ilp_scalability.tex
\begin{itemize}[leftmargin=1.8cm]
	\hypertarget{RQ2}{ }
	\item[(\textbf{RQ2})] What is the overhead of identifying and modeling dependencies?    
	\hypertarget{RQ3}{ }
	\item[(\textbf{RQ3})] How scalable is the ILP model of DARS?
	\hypertarget{RQ3.1}{ }
	\item[(\textbf{RQ3.1})] Is the ILP model scalable to large scale requirement sets?
	\hypertarget{RQ3.2}{ }
	\item[(\textbf{RQ3.2})] What is the impact of budget on runtime?
	\hypertarget{RQ3.3}{ }
	\item[(\textbf{RQ3.3})] What is the impact of precedence dependencies on runtime?
	\hypertarget{RQ3.4}{ }
	\item[(\textbf{RQ3.4})] What is the impact of value dependencies on runtime?
\end{itemize}

%% file: table_sim_scalability_design.tex
\resizebox {0.8\textwidth }{!}{
\begin{tabular}{lllllll}
	\toprule[1.5pt]
	\rowcolor{gray!30}
	\textbf{\cellcolor{black}\textcolor{white}{Simulation}} &
	\textbf{\cellcolor{black}\textcolor{white}{Size}} &
	\textbf{\cellcolor{black}\textcolor{white}{$\%$Budget}} &
	\textbf{\cellcolor{black}\textcolor{white}{VDL}} &
	\textbf{\cellcolor{black}\textcolor{white}{NVDL}} &
	\textbf{\cellcolor{black}\textcolor{white}{PDL}} &
	\textbf{\cellcolor{black}\textcolor{white}{NPDL}}
	\bigstrut\\
	\hline
	1 &
	[0,3000] &
	50 &
	0.15 &
	0.00 &
	0.02 &
	0.00
	\bigstrut\\ \rowcolor{gray!25}
	2 &
	200 &
	[0,100] &
	0.15 &
	0.00 &
	0.02 &
	0.00
	\bigstrut\\	
	3 &
	200 &
	50 &
	0.15 &
	0 &
    [0,1] &
	0.00
	\bigstrut\\ \rowcolor{gray!25}
    4 &
	200 &
	50 &
	0.15 &
	0.00 &
	0.02 &
	[0,1]
	\bigstrut\\
	5 &
	200 &
	50 &
	[0,1] &
	0.00 &
	0.02 &
	0.00
	\bigstrut\\ \rowcolor{gray!25}
	6 &
	200 &
	50 &
	0.15 &
	[0,1] &
	0.02 &
	0.00
	\bigstrut\\ 
	\bottomrule[1.5pt]
\end{tabular}
}